\documentclass[apj]{emulateapj}
\usepackage{graphicx}
\usepackage{times}
\usepackage{amsmath}
\usepackage{color}
\usepackage{graphics}
\usepackage{subfigure}
\usepackage{txfonts}
\usepackage{natbib}
\usepackage{wasysym}
\usepackage{afterpage}

\newcommand{\kps}{km\,s$^{-1}$}
\newcommand{\tA}{t_{\rm A}}

\shorttitle{Vortex and sink flows in eruptive flares}
\shortauthors{Zuccarello et al.}

\begin{document}
\title{Vortex and sink flows in eruptive flares as a model for coronal implosions }
\author{Francesco P. Zuccarello\altaffilmark{1,2}}\email{francesco.zuccarello@wis.kuleuven.be}
\author{Guillaume Aulanier\altaffilmark{2}}
\author{Jaroslav Dud\'ik\altaffilmark{3}}\email{dudik@asu.cas.cz}
\author{Pascal D\'{e}moulin\altaffilmark{2}}
\author{Brigitte Schmieder\altaffilmark{2}}
\author{Stuart A. Gilchrist\altaffilmark{4}}

\altaffiltext{1} {Centre for mathematical Plasma Astrophysics, Department of Mathematics, KU Leuven, Celestijnenlaan 200B, B-3001 Leuven, Belgium}
\altaffiltext{2}{LESIA, Observatoire de Paris, PSL Research University, CNRS, Sorbonne Universit\'{e}s, UPMC Univ. Paris 06, Univ. Paris-Diderot, Sorbonne Paris Cité, 5 place Jules Janssen, F-92195 Meudon, France}
\altaffiltext{3}{Astronomical Institute of the Academy of Sciences of the Czech Republic, Fri\v{c}ova 298, 251 65 Ond\v{r}ejov, Czech Republic}
\altaffiltext{4}{NorthWest Research Associates, 3380 Mitchell Lane, Boulder, CO 80301, USA}

\begin{abstract}

Eruptive flares are sudden releases of magnetic energy that involve many phenomena, several of which can be explained by the standard 2D  flare model and its realizations in three-dimensions. 
We analyze a three-dimensional magnetohydrodynamics simulation in the framework of this model that naturally explains the  contraction of coronal loops in the proximity of the flare sites, as well as the inflow  towards the region above the cusp-shaped loops.  
We find that two vorticity arcs located along the flanks of the erupting magnetic flux rope are generated as soon as the eruption begins. The  magnetic arcades above the flux-rope legs are then subjected to expansion, rotation or contraction depending on which part of the vortex-flow advects them. In addition to the vortices, an inward-directed magnetic pressure gradient exists in the current sheet below the magnetic flux rope. It results in the formation of a sink that is maintained by reconnection. 
We conclude that coronal loop apparent implosions observed during eruptive flares are the result of hydro-magnetic effects related to the generation of vortex- and sink-flows when a flux rope  moves in a magnetized environment.  

\end{abstract}

%

\section{Introduction}
\label{Sect:1}

Solar flares exhibit a multitude of observed dynamic phenomena arising from magnetic energy release \citep[e.g.,][]{Fle2011,Sch2015}. These phenomena constitute a set of constraints that must be naturally reproduced by any physical model that aims at being generic. 
 
In the case of eruptive flares, many observed phenomena have been interpreted in the framework of the CSHKP flare model \citep{Carmichael64,Sturrock66,Hirayama74,Kopp76}  both in its  2D version and in its  realizations in 3D \citep{Shi1995,Moore2001,Aul2012,Janvier13,Janvier15}. 

S-shaped X-ray or EUV loops seen within so-called sigmoids prior to eruptions \citep[e.g. ][]{Green09,Savcheva12a,Savcheva12b,Savcheva14,Zhao2016} as well as thick coronal loop-like structures erupting from the loci of sigmoids \citep[e.g. ][]{Cheng10,Zhang12,Patsourakos13,Cheng13,Dudik14a,Cheng14a,Cheng14b,Cheng15} have both been interpreted as typical signatures of a flux rope that is formed and eventually erupts. Observations of slipping and hot (typically 10\,MK) flare loops with their footpoints moving along J-shaped EUV ribbons have been reported \citep{Dudik14a,Dudik16,Li14,Li15,Gou16}, and narrow photospheric current ribbons have been measured along the latter \citep{Aul2012,Janvier14,Janvier16}. These observations are consistent with flux-rope related quasi-separatrix layers (QSLs), and with QSL reconnection \citep{Sav2015,Sav2016}. 

Inflows of nearly vertical loops towards each other,  below expanding CMEs have been long-since reported. These motions have been interpreted as motions towards the vertical reconnecting current sheet that is formed below the expanding flux rope \citep[e.g. ][]{Yokoyama01,Liu10b,Savage12,Takasao12,Hannah13,Zhu16}.  \cite{Su13} and \cite{Zhu16} have shown that the disappearance of the inflowing loops is followed by the appearance of hot flare loops, suggesting that these loop motions are reconnection-related inflows. These authors have also noticed that the inflows tend to accelerate towards the reconnection site with inflow velocities that range from less than 10 \kps to few hundreds of \kps \citep{Liu10b,Savage12,Su13,Zhu16}.

In the last decade several observations of contractions of closed coronal loops located at the periphery of active regions towards the flare and/or eruption site have also been reported \citep[e.g.,][]{Khan06a,Khan06b,Liu09b,Liu10a,Liu12a,Gosain12,Kallunki12,Sun12,Simoes13a,Shen14,Imada14,Kushwaha15,Petrie16,Dudik16}. These contracting loops are seen during all possible flare phases; early, impulsive, and gradual. Firstly, the most detailed observational analysis have highlighted that strong contractions (with velocities of several tens or up to a hundred \kps) occur in the impulsive phase and are usually followed by coronal dimming and loop oscillations. Secondly, it has been shown that these contractions are either preceded by, or simultaneous with the beginning of the eruption \citep[e.g.,][]{Liu12a,Shen14}.  \citet{Dudik16} have shown that the strong coronal loop contractions occur only after the onset of the fast eruption.  Thirdly, \cite{Gosain12} and \cite{Simoes13a} have shown that the onset of the contraction depends on the location of the loops with respect to the flare: loops located progressively further away from the flare contract later. Finally, \cite{Russell15} presented a common model for loop contraction and accompanying oscillations.

These coronal inflows and contractions have been interpreted as an observational confirmation of the conjecture proposed by \cite{Hudson00}, that is,  during a transient (in a low plasma-$\beta$ environment and with negligible gravity), ``the coronal field lines must contract in such a way as to reduce $E_\mathrm{mag} = \int_V (B^2/2\mu) \mathrm{d}V$.'' This conjecture is based on the premise that a decrease of $E_\mathrm{mag}$ must be accompanied by a decrease of $V$. In the case of an eruptive flare the decrease of volume must be stronger than for a confined flare of the same energy in order to compensate the eruption-related increase in $V$. In this picture, the decreasing volume should be observable as the coronal volume contracting (imploding) towards the flare site. To date, this  conjecture is the only attempt to explain the observations of contracting loops.

The aim of this paper is to show that contracting and inflowing loops naturally exist in the 3D MHD  models of eruptive flares, and to investigate which magneto-hydrodynamics processes generate the observed loop's behavior. To this purpose we analyze the relation between the magnetic forces and fluid dynamics effects in a zero-$\beta$ three-dimensional MHD simulation of a torus unstable erupting flux rope. 

The plan of our paper is as follows. In the next Section we outline the evolution of different sets of field lines present in the numerical simulation.  Section~\ref{Sec:T-Vortex} is devoted to the analysis of the 3D distribution of the flows that are generated during the outward propagation of the magnetic flux rope, while in Section~\ref{Sec:T-Inflow} we describe the flows towards the current sheet. Finally, in Section~\ref{Sect:Discussion} we discuss our findings and their relation with the implosion conjecture. 

\section{Model}
\label{Sect:Model}


\subsection{Numerical simulation}
\label{Sec:Numerical}

The dynamics of the magnetic field during the occurrence of erupting flares is modeled using the OHM-MPI code \citep{Zuc2015}. In particular, for this paper we analyze the later stages, i.e., from the onset of the torus instability onward, of the simulation labeled as `Run~D2' in \cite{Zuc2015}. In this simulation a magnetic flux rope is formed through flux cancellation at the polarity inversion line (PIL) of a previously sheared magnetic arcade.  The details of the numerical setup have been extensively discussed in Section~2 and Section~4.5 of \cite{Zuc2015}. 

However, we  retain the important information that the flux rope becomes torus unstable at around $t_{\text{crit}}=165~\tA$, and undergoes a full eruption. During this phase the pre-eruptive photospheric drivers are reset to zero, i.e., the system is allowed to evolve only under the effect of the Lorentz forces already present in the system.

The simulation ends at $t=244~\tA$. As already discussed in \cite{Zuc2015} the simulation is performed on a non-uniform grid that expands starting from the center of the domain. While the flux rope moves upwards and sideways also the flare current sheet eventually reaches a location in the domain where the grid resolution is not sufficient to properly resolve the local gradients, so a numerical instability sets in eventually halting the simulation. While this has no impact on the simulation apart from the last Alfv\'{e}n time, it prevents to follow the longer-term development of the simulated eruption.

The evolution of the coronal magnetic field surrounding the flux rope  as driven by the eruption constitutes the subject of this paper. This dynamics is only related to the presence of moving current-carrying magnetic fields and it is, in principle, independent from the exact trigger of the eruption.

\begin{figure*}
\begin{center}
\includegraphics[width=0.99\textwidth,viewport= 50 0 2193 1154,clip]{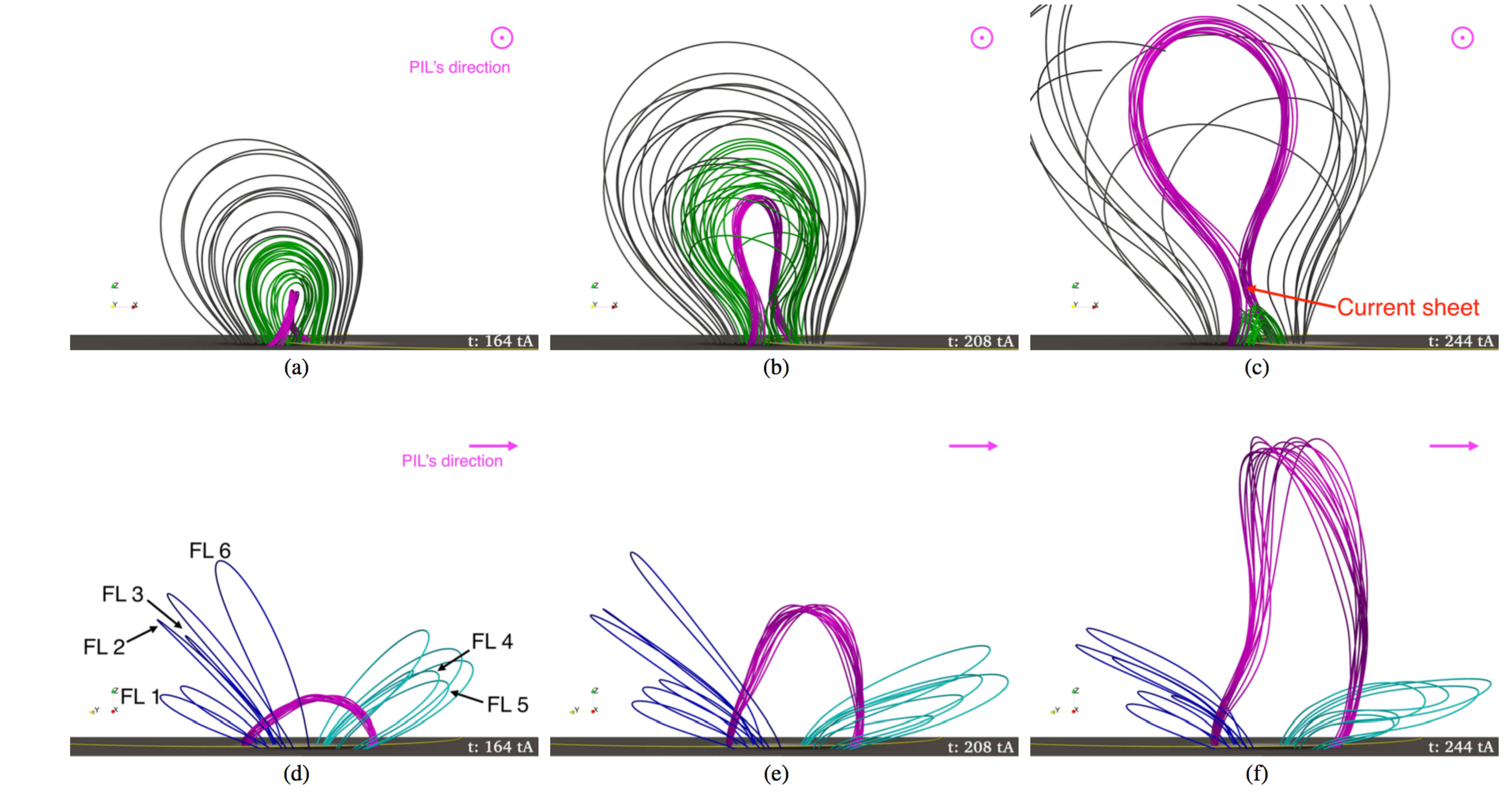}
%
\caption{Projected views of the evolution of the system as seen from planes almost perpendicular (\textit{a,b,c}) and parallel (\textit{d,e,f}) to the central portion of the PIL and flux rope axis. The magenta field lines highlight the erupting magnetic flux rope. The overall direction of the PIL and of the axial field of the flux rope is indicated by the pink dotted circle (magnetic field vector exiting the plane) in the top panels and by the pink arrow in the bottom panels. The green/gray field lines highlight the portion of the overlying field that is rooted in a region comprised between  the PIL and the strongest field (the center of the polarities), while the blue/cyan field lines are rooted in the polarities' periphery. The red arrow in  panel (c) indicates the position of the flare current sheet below the erupting flux rope (see also Figure~\ref{Fig:Div_u_3D}c).  The numbers in panel (d) are used to identify selected field lines which dynamics is described in Section~\ref{Sect:Model}. In each row field lines with the same photospheric anchorage are shown.
(An animation of this figure, Movie~1, is available in the online version of the article.)  }
\label{Fig:Evol}
\end{center}
\end{figure*}
%

\begin{figure*}
\begin{center}
\includegraphics[width=.95\textwidth]{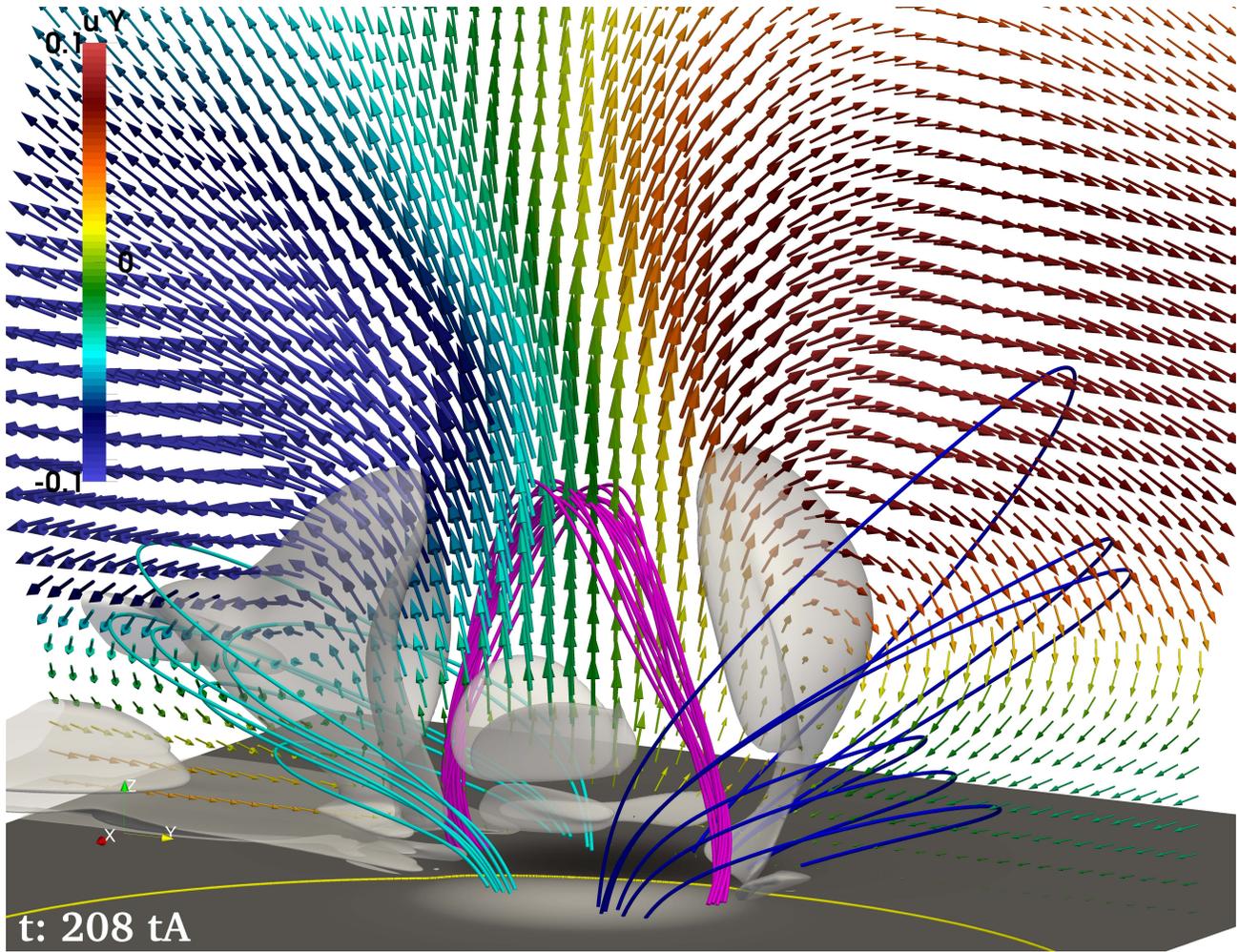}
\caption{Three-dimensional view of the erupting magnetic flux rope (magenta field lines) and selected (cyan/blue) peripheral magnetic field lines. The arrows represents the 3D velocity field along the $yz$-plane passing through $x=-0.75$. The arrows are color-coded with the $y$-component of the velocity. The semi-transparent grey surface is an iso-contour of the absolute value of the component of the flow vorticity that is perpendicular to the seed plane for the 3D velocity vector, i.e, an iso-contour of $\vert[\nabla \times \mathbf{u}]_x\vert = 0.05$. 
(An animation of this figure, Movie~2, is available in the online version of the article.)}
\label{Fig:3D-Rot_u}
\end{center}
\end{figure*}


\subsection{Flux rope and the neighboring coronal magnetic field}
\label{Sec:T-Loops}

As a consequence of the outward motion of the magnetic flux rope (magenta field lines in Figure~\ref{Fig:Evol}), the top part of the magnetic envelope that surrounds the flux rope expands upward and outward in all the directions, while the bottom part of the legs are seen to converge towards the current sheet that is formed below the erupting flux rope and above the PIL (see Figure~\ref{Fig:Evol}, top-panels and  accompanying Movie~1). This inflow of the legs of the overlying field towards the current sheet is more visible following the green field lines in Figures \ref{Fig:Evol}(a) and \ref{Fig:Evol}(b), i.e., between $t=[164~\tA, 208~\tA]$, and the gray field lines in Figures \ref{Fig:Evol}(b) and \ref{Fig:Evol}(c), i.e., between $t=[208~\tA,244~\tA]$, when more and more field lines are advected towards the current sheet.  

Figure~\ref{Fig:Evol} also shows the change of connectivity of the green field lines. While the flux rope propagates upward overlying magnetic field lines are advected towards the current sheet, where they eventually reconnect. At time $t=244~\tA$, all the green field lines that originally were part of the overlying magnetic field, have now reconnected at the current sheet and close below it, displaying the typical cusp shaped structure often observed in post flare loops \citep[e.g.,][]{Yokoyama01,Liu10b}. 

Therefore, the simulation predicts that when the eruption is seen from a plane that is (almost) perpendicular to the strongest portion of the PIL, the side of the magnetic envelope, which surrounds the flux rope, initially moves apart to let the erupting flux rope pass, but eventually the  legs of the envelope are seen to move back toward the current sheet that is formed below the erupting field.

The evolution of the system as seen from the side is shown in Figure~\ref{Fig:Evol} (bottom-panels) and  accompanying Movie~1. 
The blue/cyan magnetic field lines highlight the part of the external background magnetic field that is rooted on the weaker peripheral magnetic field of the active region. 
As soon as the flux rope begins to expands, the cyan/blue field lines  start to move as well. These magnetic field lines are seen to undergo an expansion and a decrease in altitude before contracting towards the flanks of the flux rope (the flanks are defined as the external sides of the flux rope's legs). 

The analysis of Figure~\ref{Fig:Evol} (bottom-panels) and Movie~1 shows that while the cyan/blue field lines all display a similar behavior, the importance of the different kind of motions is different for different field lines. 
In particular:  Field Line~1 (FL~1) shows a minor expansion and rotation before contracting;  FL~2 shows a more pronounced rotation and expansion and starts to contract only towards the end of the simulation;  FL~3 shows a contraction that is much more important than its expansion. An analysis of Movie~1 shows that FL~3 is seen to contract while FL~1 and FL~2 are still expanding. In particular,  FL~3 is the first that starts to contract, followed by FL~1 ---while FL~2 is still expanding--- and finally FL~2 starts to contract as well. An analysis of the cyan field lines reveals a similar behavior. In particular, the neighboring FL~4 and FL~5 both show an initial expansion. However, after about $t=192~\tA$ FL~5 inverts its behavior and starts to contract, while FL~4 continues to expand. 

Hereafter, we investigate the dynamic of the system to determine the physical mechanism that is responsible for the observed inflow  and expansion/contraction motions.

\section{Vortex flows}
\label{Sec:T-Vortex}

In this Section we describe the generation and propagation of vortex flows during the eruption of the magnetic flux rope. 

\subsection{The presence of vortex flows}

Analysis of the plasma flow field around the selected moving field lines (as described in Section~\ref{Sec:T-Loops}) reveals the presence of rotating flows in the form of two large-scale vortices (see arrows in Figure~\ref{Fig:3D-Rot_u} and accompanying Movie~2). 

These vortex-flows are located above the PIL and at each flank of the flux rope. The plasma velocities inside them correspond to a fraction of the upward-directed velocity around the central part of the erupting flux rope. 

Each of these two large-scale vortices is associated with an extended source of vorticity located at its center, that is most visible when only its component orthogonal to the flow field is plotted (see the semi-transparent gray surface in Figure~\ref{Fig:3D-Rot_u}).

Movie~2 shows that the general circulation of these flows is established very early in the eruption. This can be seen in the bottom-right part of the domain (around the dark blue loops), in the time-interval [$164~\tA ; 180~\tA$]. There, the velocity arrows progressively  change direction, from an initial overall expansion towards a vortex-like circulation, firstly close to the flux rope and later away from it.  This evolution and its duration roughly correspond to the transit of an Alfv\'{e}n wavefront from the location of the flux rope (around $y=0$) towards the boundaries of the numerical domain (at $y=\pm10$ and $z=30$). The same behavior can be seen on the other side of the flux rope (around the cyan loops), but there it is not as straightforward, primarily because the deflection  of the erupting flux rope in this direction alters the velocity pattern \citep[see][ for more detail on the flux rope's deflection]{Zuc2015}. 

At later times, Figure~\ref{Fig:3D-Rot_u} at $t=208~\tA$ and Movie~2 show that the vortices extend in size both horizontally and vertically. Firstly, the associated flow pattern eventually fills the whole domain, including up to altitudes far above the apex of the erupting flux rope. Secondly, the rotating velocities increase in magnitude, co-temporally with the accelerating flux rope, which in our model results from the development of the torus instability. Thirdly, the  vorticity concentrations move away from the erupting flux rope, and they rise in altitude. Finally, as the boundary effect of the photosphere diminishes with time, the flow pattern around each vorticity concentration becomes more and more circular with time. 

In spite of the associated smooth rotational pattern, these quickly-established and slowly-strengthening vortex flows only lead to an incomplete revolution of plasma and loops around their centers. The analysis of the modeled (dark blue and cyan) loop dynamics shows that their rotation is not greater than $\pi/4$ during the time-interval [$164~\tA ; 244~\tA$]. Also, the  field line  motion along the rotational flow is continuous from the beginning of the eruption up to the end of the calculation.  So, the  expansion and contraction of the field lines is just a manifestation of their location within the vortex flow at a given time. 

Finally, there are two other concentrations of vorticity in the domain (Figure~\ref{Fig:3D-Rot_u}). One region is located below the erupting flux rope associated with the flare current sheet, due to the reconnection-driven deflection of the flows around it (see Section~\ref{Sec:T-Inflow}). Another elongated region is located close to the photosphere above the PIL, due to a boundary layer that is created by the compression of the vortex flows by the asymmetric eruption.

\subsection{Physical origin of the vortices}

The sudden formation of the vortex-flows and the continuity of these flows are reminiscent of the same behavior as observed for the generation of vortex rings from a vortex cannon, or during a thermal explosion, and as observed for the generation of vortex pairs around a fluid element that moves within a non-moving environment as modeled by  \cite{Belmont2013}. 

This surprising phenomenological similarity may be challenged a-priori for the four coupled reasons. 
First, the modeled system is zero-$\beta$ and therefore does not involve any plasma pressure.
Second, the magnetic and flow field geometries involve complex couplings between the magnetic pressure force $-\nabla (B^2/2)$, the tension force $(\mathbf{B}\cdot \nabla)\mathbf{B}$, and the non-linear term $\rho(\mathbf{u}\cdot \nabla)\mathbf{u}$ in the momentum equation that governs the acceleration of the frozen-in plasma.
Third, the magnetic forces could resist the vortex flows. 
Finally, the vortices strengthen in time, which does not occur in vortex-cannon experiments where the vortex are readily generated by the sudden outflows. 

These arguments can actually be subjected to an analysis of the forces at work in the simulation. Such an analysis is provided in the following subsections. 

\begin{figure*}
\begin{center}
\subfigure{
\includegraphics[width=.75\textwidth]{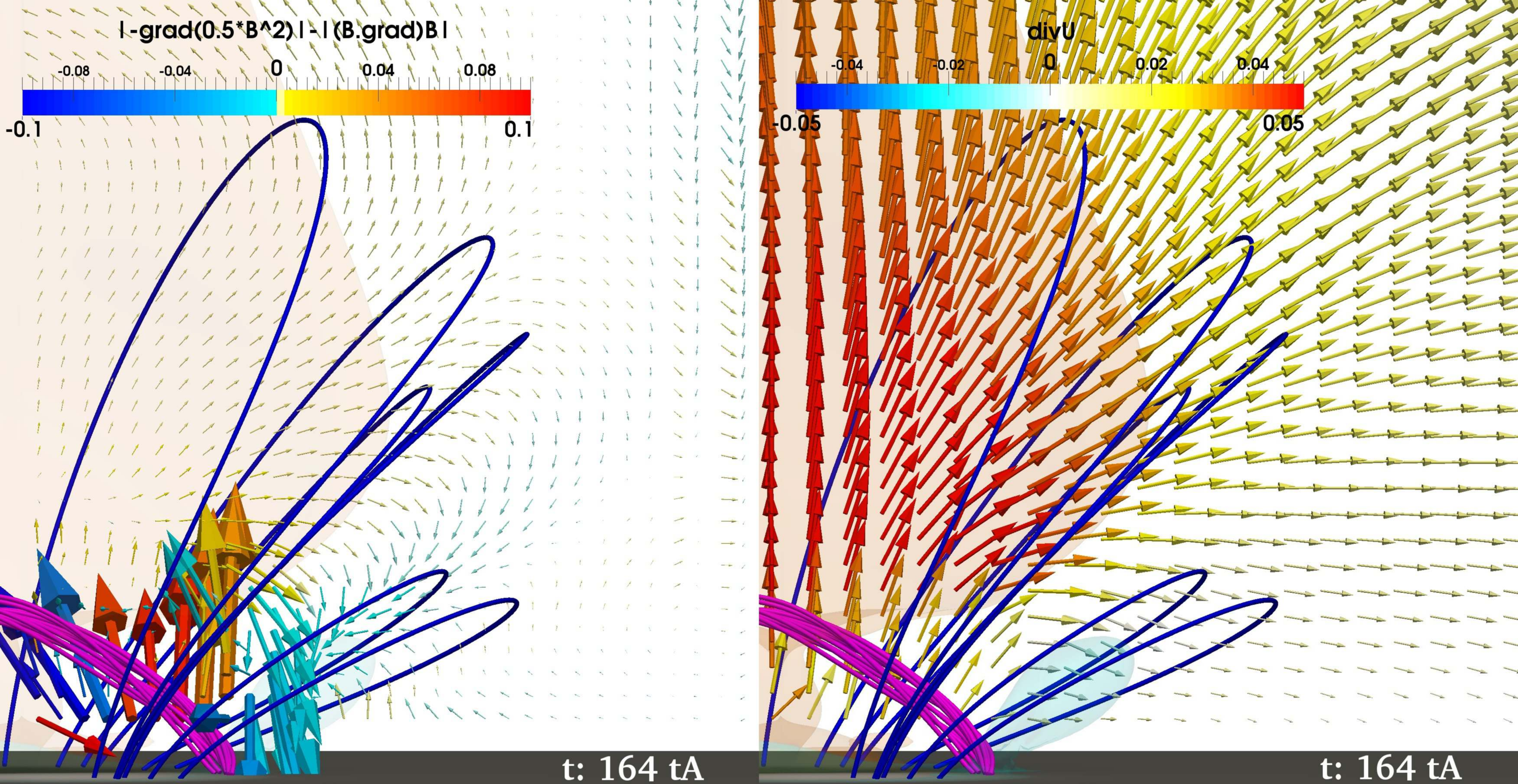}\label{Fig:Force_early1}} 
\subfigure{
\includegraphics[width=.75\textwidth]{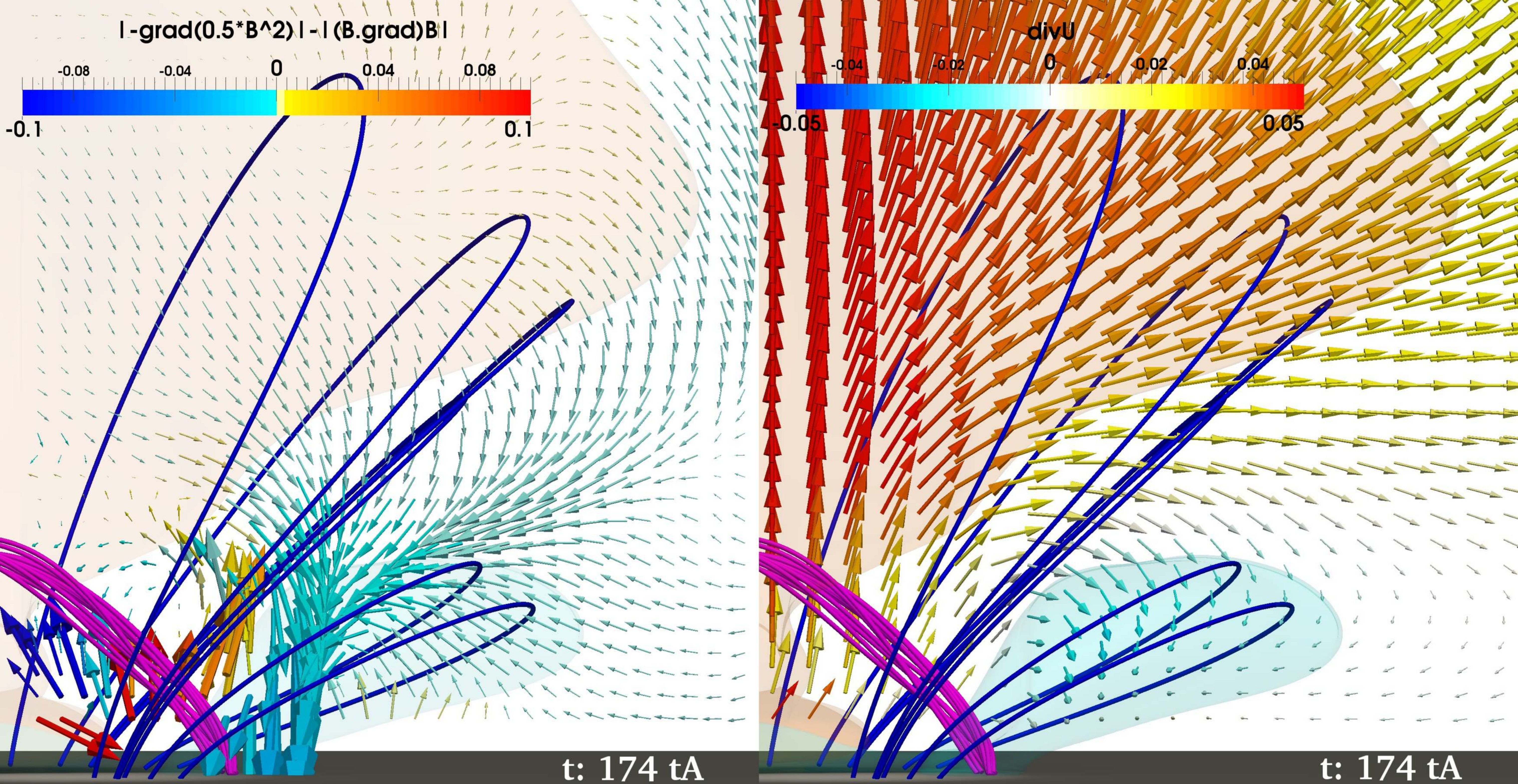}\label{Fig:Force_early2}} 
\subfigure{
\includegraphics[width=.75\textwidth]{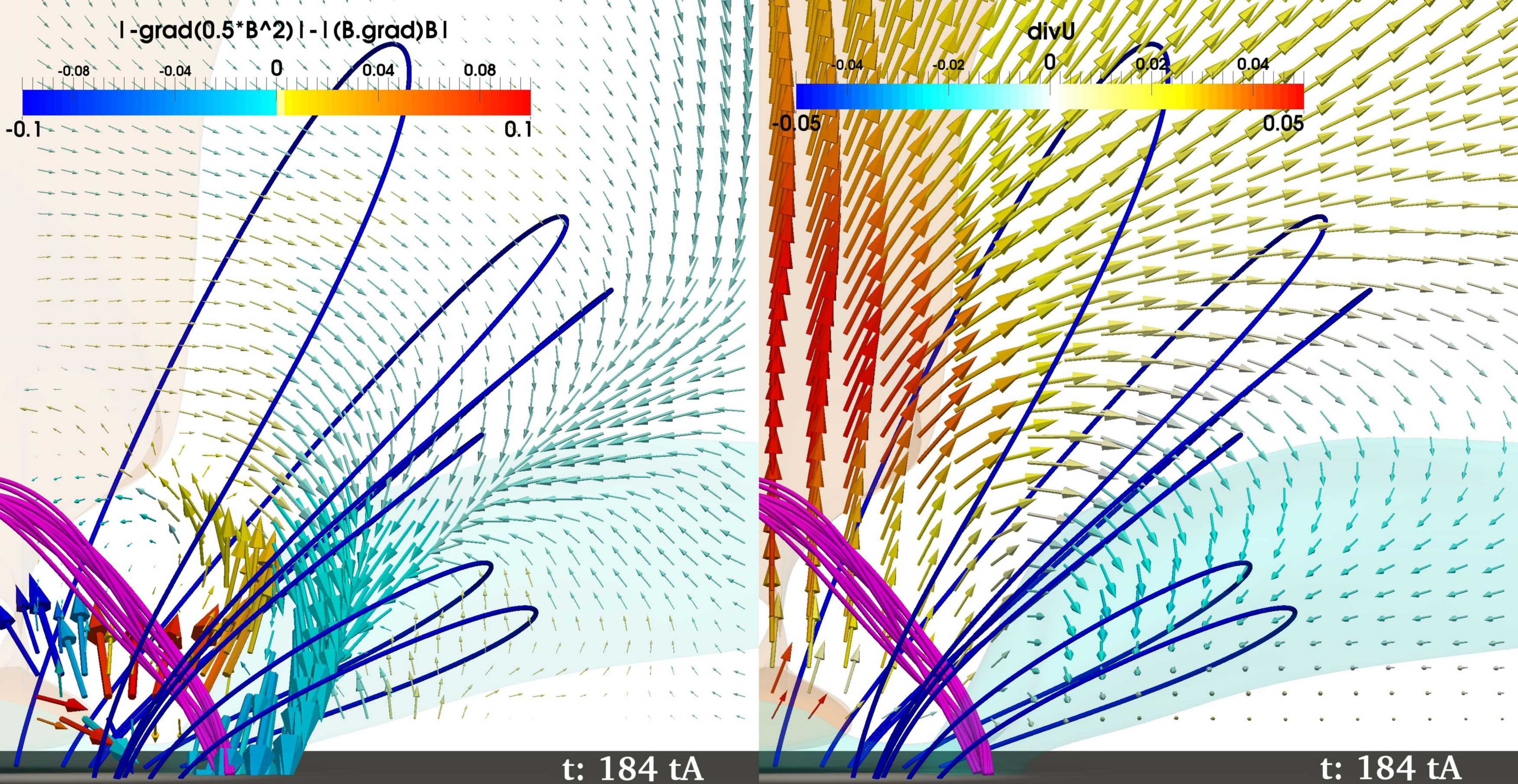}\label{Fig:Force_early3}} 
\caption{Snapshots of the simulation up to 20 $~\tA$ after the onset of the torus instability. The panels show a portion of the erupting flux rope (magenta field lines) and selected side magnetic field lines (blue). The orange (cyan) surface is an iso-surface of $\nabla \cdot \mathbf{u}$ = 0.04 (-0.02). The arrows on the left-panels are the Lorentz force normalized to the density and color coded with $| - \nabla(B^2/2)| - | (\mathbf{B} \cdot \nabla ) \mathbf{B}| $, while the arrows on the right panels are the velocity vectors color coded with $\nabla \cdot \mathbf{u}$. In all panels the arrows are traced starting from a $yz$-plane passing through x=-0.75.  
(An animation of this figure, Movie~3, is available in the online version of the article.)  }
\label{Fig:Force_early}
\end{center}
\end{figure*}

\begin{figure*}
\begin{center}
\subfigure{
\includegraphics[width=.75\textwidth]{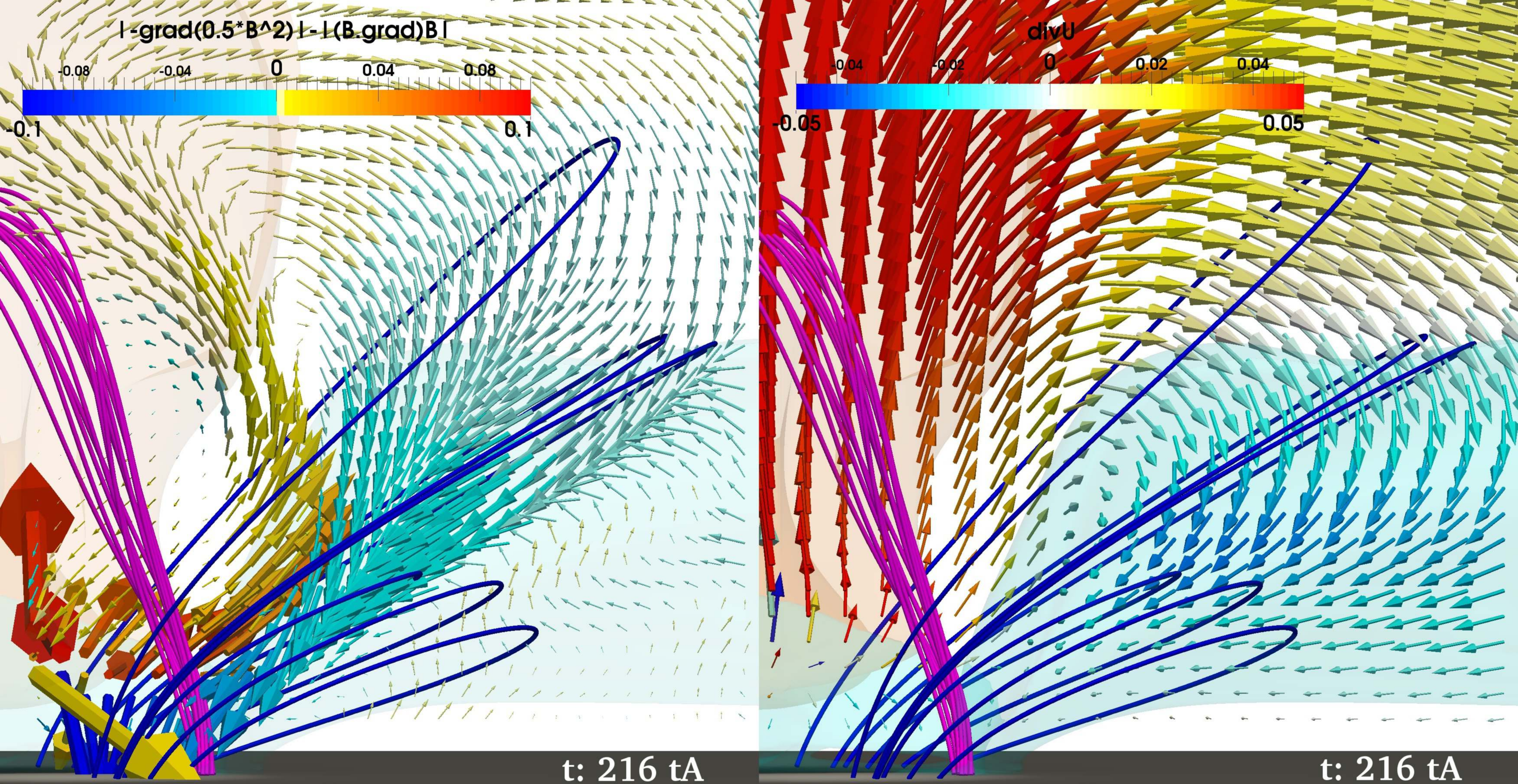}\label{Fig:Force_late1}} 
\subfigure{
\includegraphics[width=.75\textwidth]{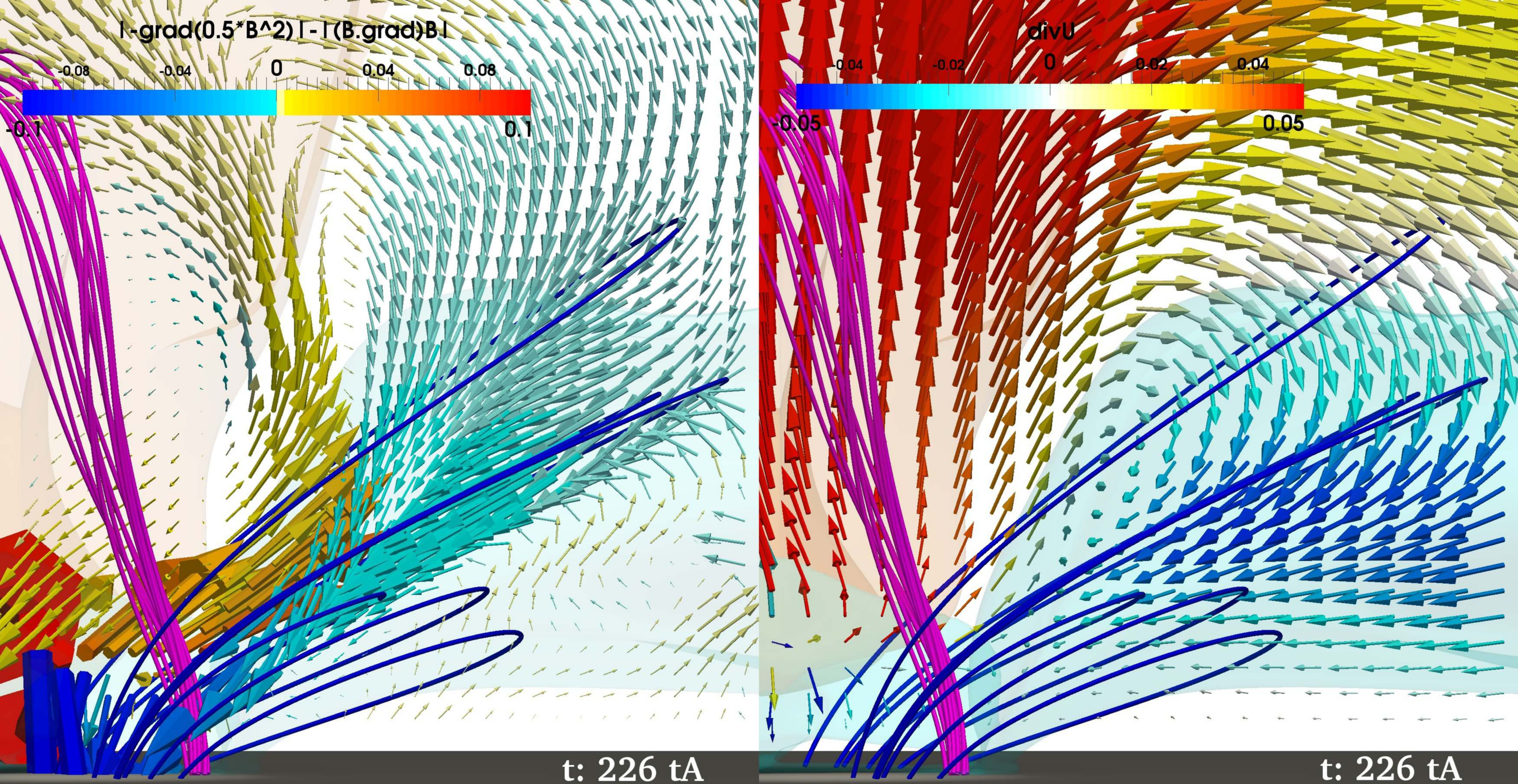}\label{Fig:Force_late2}} 
\subfigure{
\includegraphics[width=.75\textwidth]{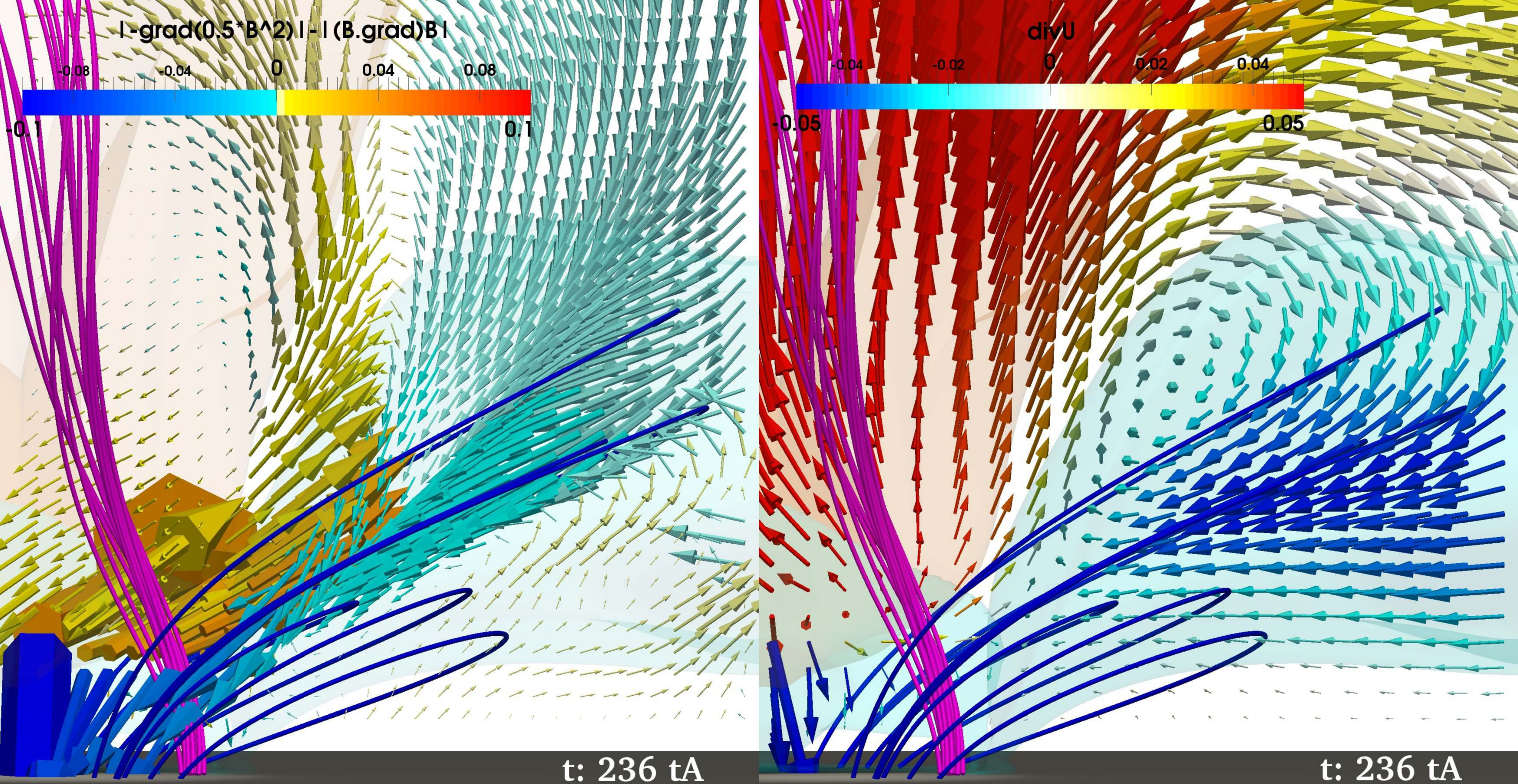}\label{Fig:Force_late3}} 
\caption{Same as Figure~\ref{Fig:Force_early}, but for a later stage of the simulation. 
(An animation of this figure, Movie~3, is available in the online version of the article.)  }
\label{Fig:Force_late}
\end{center}
\end{figure*}

\subsubsection{Compressible and solenoidal flow components}
               
\cite{Belmont2013} have shown that in 2.5D an initially-confined low Mach number velocity pulse results in the generation of two vortices on its sides (see their Figures.~9.7-9.8 in pages 367-368). These authors have shown that two pressure wave fronts are suddenly generated and propagates into the domain. The important finding  is that the boundary of the region filled with the vortices coincides with the pressure wave front at all times during the propagation of this wavefront. They concluded that this is an evidence that vortices are indeed accelerated by the advancing pressure wave fronts. 
 
The same authors (R. Grappin, private communication) have shown that the initial flow can actually be decomposed into a solenoidal flow ($\nabla \cdot \mathbf{u} =0$) and a compressible flow. The solenoidal flow has
two vortex flows centered on two $\nabla \times \mathbf{u} = \boldsymbol{\omega}$ concentrations aside of the moving fluid.  The compressible flow has two components related with a $\nabla \cdot \mathbf{u} <0$ sink located ahead of the moving fluid element and a $\nabla \cdot \mathbf{u} >0$ source behind it. The compressible component of the flow is evacuated at the characteristic (sonic or magnetosonic) speed away from the slowly moving fluid element (hence the aforementioned pressure wave), while the solenoidal component is left behind with its two vortices slowly moving the fluid in the whole domain (this component has no associate wave as it is incompressible: $\nabla \cdot \mathbf{u} =0$).

\subsubsection{Vortex early formation}
 \label{Vortex-early}

Figure~\ref{Fig:Force_early} shows the forces, velocities and $\nabla \cdot \mathbf{u}$ contours on one side of the erupting flux rope during the early formation phase of the vortex-flows. 

The link with the vortex generation in \cite{Belmont2013} is not straightforward since the analyzed systems are different. There are at least four main differences between our model and the above hydrodynamic case.   
 First, the length of the wave packet is much larger in our model than in the simple hydrodynamic case (probably because the flux rope is continuously accelerated, unlike the fluid element in the hydrodynamic case).
 Second, the wave generates a magnetic tension force in our model while it generates a thermal pressure gradient in the hydrodynamic case. 
 Third, the vortex flows are generated in a fluid that is initially at rest in the hydrodynamic case while in our model the forces first need to deflect pre-existing coronal flows that are present because of  the preceding flux rope formation phase. 
 Fourth, the erupting magnetic configuration is anchored to the photospheric boundary.

A global expansion, $\nabla \cdot \mathbf{u}>0$, of the plasma is present in front of the erupting flux rope, already before the eruption starts. This presumably masks part of the  $\nabla \cdot \mathbf{u} <0$  compressible component formed in front of the flux rope (i.e. the moving fluid of Belmont et al. 2013), which in our simulation only manifests itself as a sudden diminution of the extended $\nabla \cdot \mathbf{u}>0$ region. However, the  $\nabla \cdot \mathbf{u} <0$ compressible component is directly visible on the flanks of the erupting flux rope, where the global expansion is weaker.  As in the 2.5 D simulation of \cite{Belmont2013}, we interpret the formation of the vortices by the solenoidal flow component left behind the propagating compressible flow component.
Indeed, the $\nabla \cdot \mathbf{u}<0$ region (resp. the sudden decrease of $\nabla \cdot \mathbf{u}>0$) is initially located on the side of the flux rope's flank (resp. on the top of the flux rope), then it travels at the characteristic Alfv\'{e}n speed. It generates magnetic tension force and accelerates the vortex-flow. In front of the flux rope, the tension force associated with the passage of the compressible wave is then replaced by the pressure gradient force associated with the expansion of the flux rope (see Section~\ref{Vortex-long} and left columns of Figures~\ref{Fig:Force_early} and \ref{Fig:Force_late}).

Another interesting behavior is that the $\nabla \cdot \mathbf{u} <0$ component amplifies a relatively strong and narrow region of magnetic tension, which is located aside of the flux rope's flanks. This tension force originally points downwards towards the photosphere. While the wave propagates, it extends farther and farther away from the flux rope, and it induces a component of the force that points towards the edge of the flux rope's flank. Careful inspection of Figure \ref{Fig:Force_early} shows that this force does not accelerate the flow towards the flux rope. Instead, this force is orthogonal to the velocities of the vortex. It is therefore arguable that this centripetal tension force balances the centrifugal acceleration of the vortex flow, and thus maintains the vortex integrity, just like thermal pressure radial gradients do in the hydrodynamic case. The difference is that in the MHD model, the centrifugal effect of the flow does not lead to a pressure deficit. It rather tends to stretch field lines. So a curvature-related tension force at each field line apex is generated, until it halts the centrifugal growth of the vortex. 

Finally, Movie~3 shows that the arcades move in the direction of the flow, so their internal magnetic forces are obviously not strong enough to brake the vortex. The reason is probably that, since the velocity gradients along streamlines are small, and since the magnetic field locally points orthogonally to the plane of the flows, no significant magnetic pressure gradient is generated along the streamlines \citep[as also noted by][ in their 2.5D simulations]{Belmont2013}. An exception to this is close to the photospheric plane, which acts as a wall that confines the vortex. This 
boundary effect is probably responsible for the asymmetry between the top and the bottom of the vortex at these early times when the vortex has still not moved far above the photosphere. 

In summary, the key result of this analysis is that, during the early stages of the eruption, the vortices are formed neither by a stationary magnetic pressure increase ahead of the central part of the flux rope, nor by a stationary magnetic pressure deficit in the wake of the flux rope. The vortices are rather formed by a fairly broad propagating compressible wave packet that generates magnetic forces in its passage. The vortices eventually stop expanding because a strong tension in the magnetic arcades works against their centrifugal effect. 

\begin{figure*}[t!]
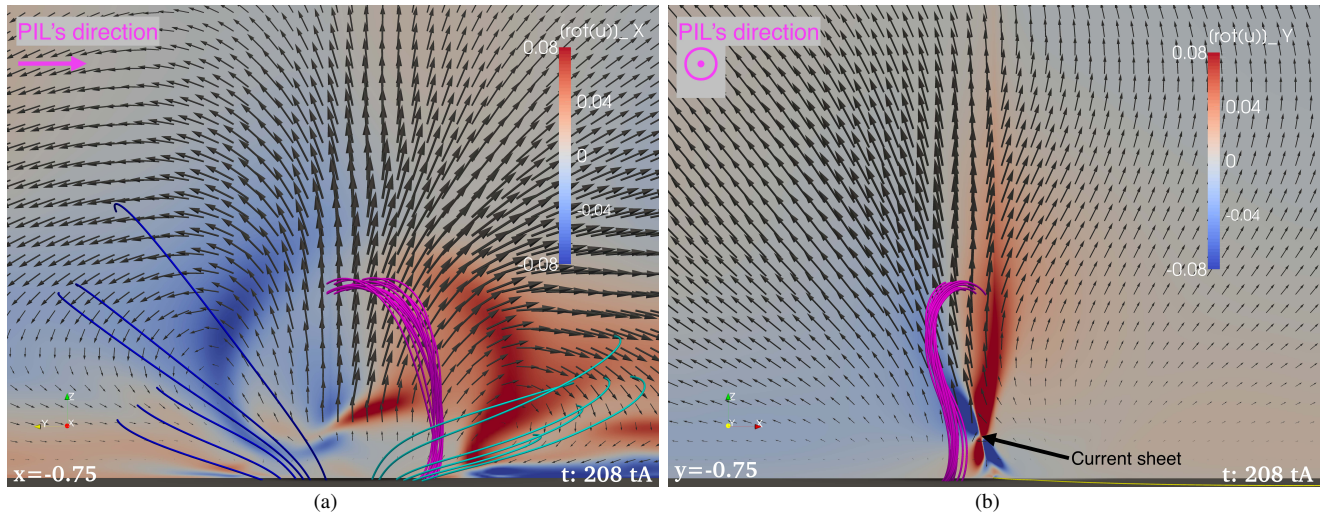

\begin{center}
\subfigure[]{
\includegraphics[width=.48\textwidth]{2D_rot_u_x.pdf}\label{Fig:Rot_u-xzview1}}
\subfigure[]{
\includegraphics[width=.48\textwidth]{2D_rot_u_y_new.pdf}\label{Fig:Rot_u-xzview2}}
\caption{Projected 2D views of the system along the $yz$-plane ($a$) and $xz$-plane ($b$) passing through $x= -0.75$ and $y=-0.75$, respectively. Only the portion  of the erupting magnetic flux rope (magenta field lines) and of the peripheral magnetic field lines (cyan/blue, only for panel ($a$)) that are in front of the opaque planes is shown.  In each panel, the color scale indicate the magnitude of the component of the flow vorticity perpendicular to the 2D projected view, that is  $[\nabla \times \mathbf{u}]_{x}$ for the ($a$) panel and $[\nabla \times \mathbf{u}]_{y}$ for ($b$) panel. Red/blue indicates vorticity component into/out-of the plane, respectively.  The arrows highlight the velocity field along the respective planes. } 
\label{Fig:Rot_u-xzview}
\end{center}
\end{figure*}

 \subsubsection{Vortex long-term development}
 \label{Vortex-long}

During the later phase, Figure \ref{Fig:Force_late} shows how the vortices are amplified and how they move, while the erupting flux rope is globally accelerated upwards and starts to expand in all directions. Both motions lead to an increase in magnetic pressure in front of the erupting flux rope.  Then, the magnetic pressure gradient eventually replaces the formed magnetic tension force that had been induced by the initial Alfv\'{e}n wave. During this phase this magnetic pressure gradient accelerates the upper half of the vortex. And since the over-pressure  is continuously strengthened because of the expansion of the  erupting flux rope, the vortex flow is continuously amplified. 

As a response to this strengthening of the vortex flows, the magnetic tension also increases in the lower half of the vortex, so as to balance the increasing centrifugal force. And the area in which the tension acts gradually moves upward and away from the flux rope, along with the vortex itself. 

This vortex displacement away from the flux rope is apparently due to another noticeable difference with the early stage of the vortex formation. This difference is that a strong pressure gradient eventually develops on the external side of the flux rope, pointing towards the vortex center. It is most likely that the latter is a reaction of the tendency of the centrifugally growing vortex to push against the flux rope's flank, and therefore to compress it. Eventually, this pressure gradient balances the centrifugal acceleration on this side of the vortex, just like the magnetic tension does on its other side, and it eventually pushes the vortex away from the flux rope. 

In summary, the motions of the arcades towards the flux rope are merely due to the early-generated and later-amplified vortices, and is completely independent of the flare reconnection and of the diminishing magnetic pressure behind the rope.

\begin{figure}
\begin{center}
\includegraphics[width=.48\textwidth]{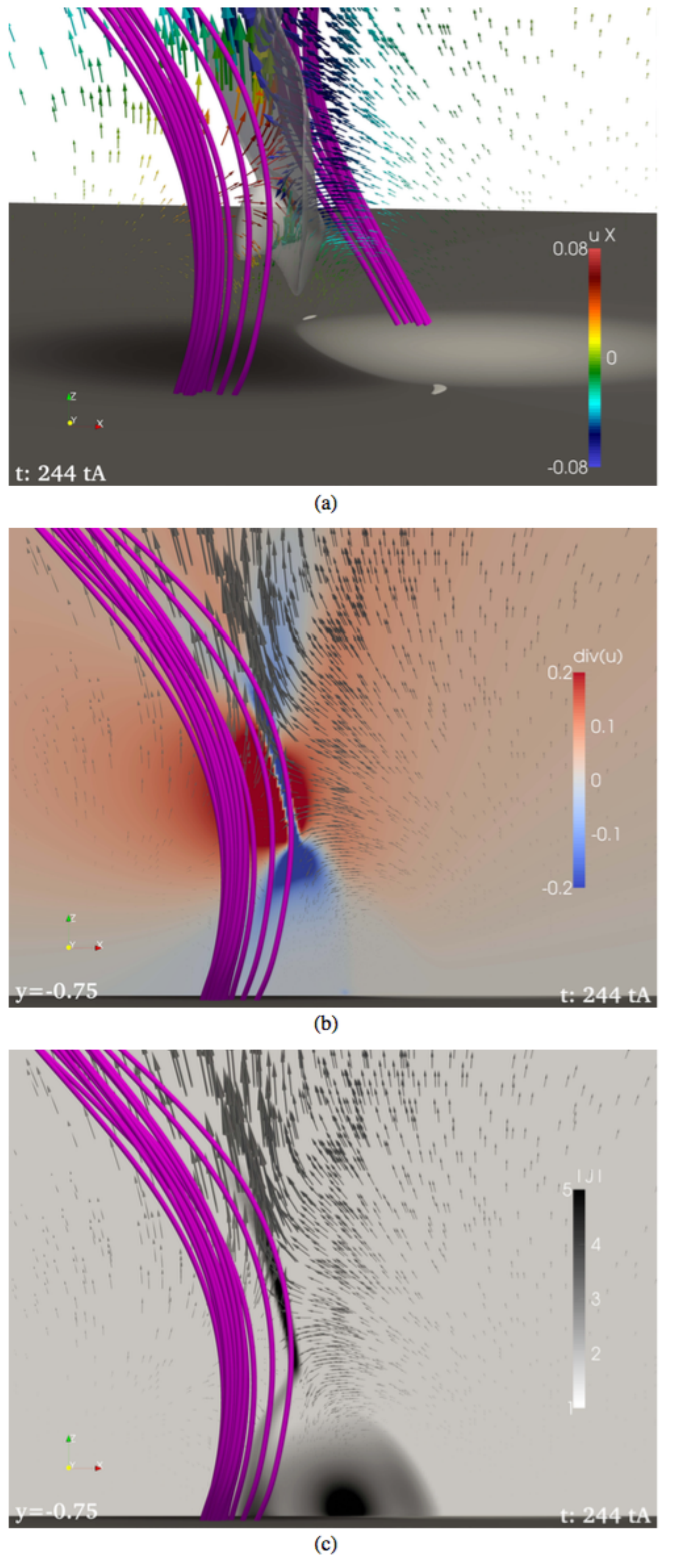}
%
\caption{Zoom into the current sheet and of the flows around it. 
($a$) 3D view of the system showing an iso-contour of $\nabla \cdot \mathbf{u} = -0.15$ with a semi-transparent grey surface. ($b$) 2D projected view showing the distribution of $\nabla \cdot \mathbf{u}$ along the $xz$-plane passing through $y=-0.75$.  ($c$) Same view as (b), but showing $|J|$.
In all panels the arrows highlight the velocity field along the xz-plane passing through $y=-0.75$. They are displayed using a random distribution, color-coded with $u_x$ for panel ($a$) and in grey for panel ($b,c$).
In all panels the magenta lines outline the erupting flux rope.
 (An animation of panel (a), Movie~4, is available in the online version of the article.) 
 }
\label{Fig:Div_u_3D}
\end{center}
\end{figure}
%

\begin{figure*}
\begin{center}
\includegraphics[width=.99\textwidth,viewport= 143 37 677 499,clip]{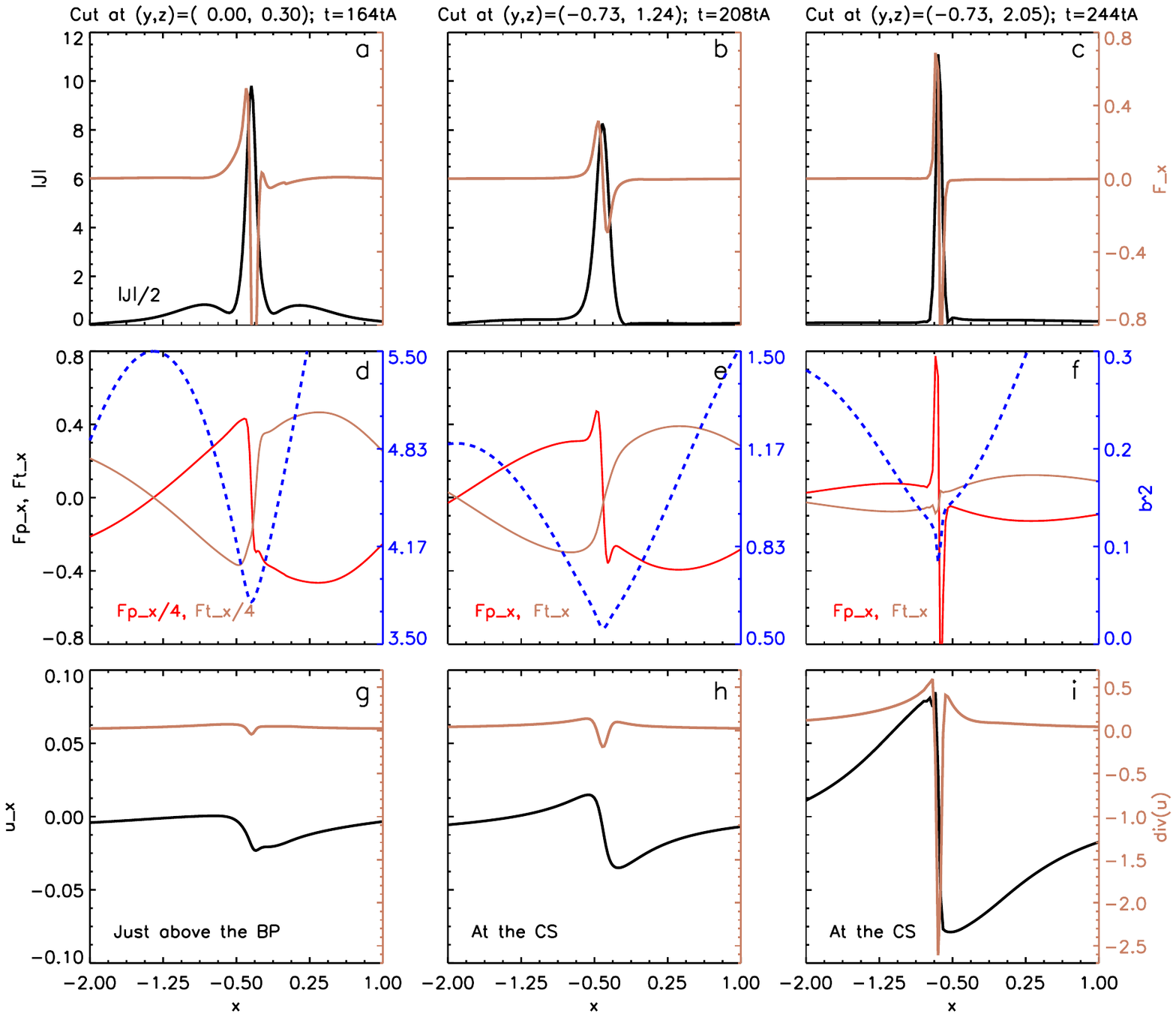}
\caption{1D plots across the current sheet (CS) or above the bald-patches (BP, left-column) showing the evolution of different relevant quantities as function of $x$ at time $t=164, 208$ and $244~\tA$ (left, middle and right columns respectively). The top row shows the evolution of $|J|$ (black) and of the $x$-component of the Lorentz force (brown), while the middle row shows $- \partial_x (B^2/2)$ (red), $(\mathbf{B} \cdot \nabla)~B_x $ (brown) and $B^2$ (dashed blue). The bottom row shows $u_x$ (black) and  $\nabla \cdot \mathbf{u}$ (brown). For better clarity in the visualization at time $t=164~\tA$, when the flux rope is at lower heights the magnetic pressure and tension have been divided by a factor 4, while the current density has been divided by a factor 2.}
\label{Fig:1D-Rot_u}
\end{center}
\end{figure*}
\begin{figure}
\begin{center}
\includegraphics[width=.49\textwidth, viewport= 51 44 1301 899]{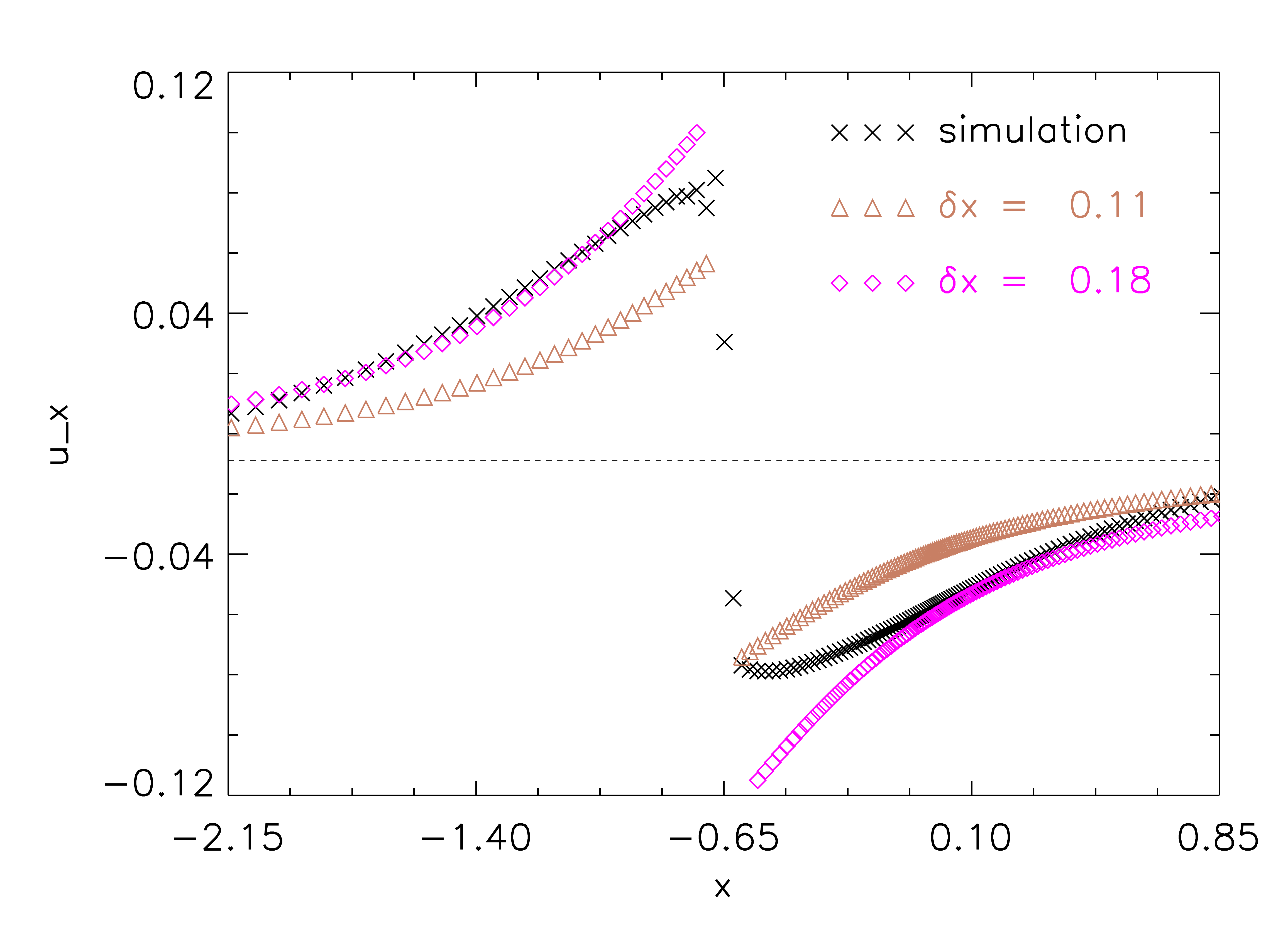}
\caption{1D cut along $y=-0.73, z=2.05$ of $u_x$ at time $t=244~\tA$ (black $x$) as well as the analytic flow for two different values $\delta x$. }
\label{Fig:1D_analyt}
\end{center}
\end{figure}


\begin{figure*}
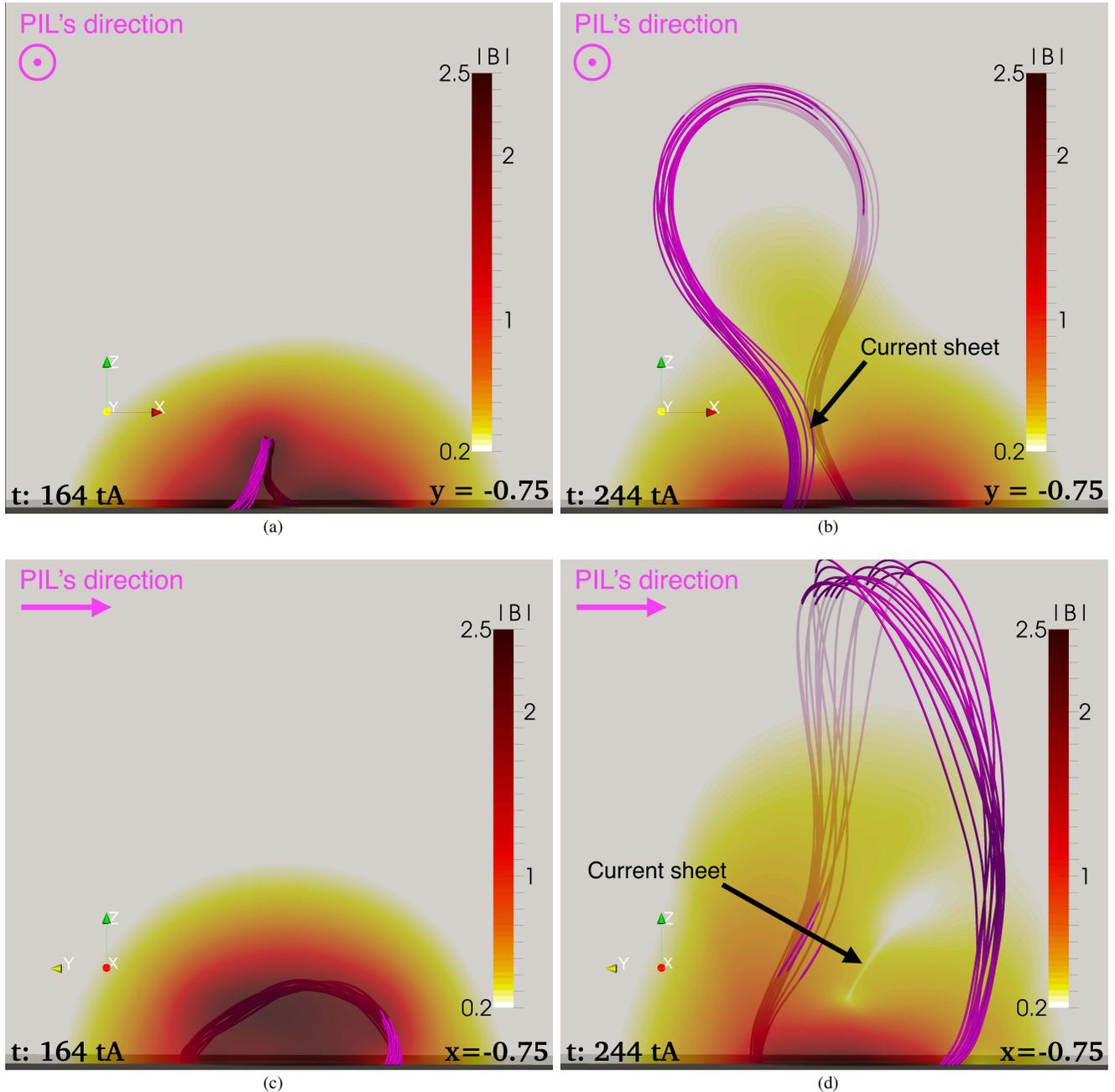

\begin{center}
\subfigure[]{
\includegraphics[width=.48\textwidth,viewport= 3073 0 6120 2850,clip]{b_mag_164ta_new.pdf}\label{Fig:B_mag1}} 
\subfigure[]{
\includegraphics[width=.48\textwidth,viewport= 3073 0 6120 2850,clip]{b_mag_244ta_new.pdf}\label{Fig:B_mag2}}
\\
\subfigure[]{
\includegraphics[width=.48\textwidth,viewport= 0 0 3050 2850,clip]{b_mag_164ta_new.pdf}\label{Fig:B_mag3}}
\subfigure[]{
\includegraphics[width=.48\textwidth,viewport= 0 0 3050 2850,clip]{b_mag_244ta_new.pdf}\label{Fig:B_mag4}}
\caption{Projected 2D views \textbf{of the initial (left column) and final (right column) distribution of magnetic field} along the $xz$-plane ($a,~b$) and $yz$-plane ($c,~d$) passing through $y= -0.75$ and $x=-0.75$, respectively. In each panel, the color scale indicate the magnitude of the magnetic field in that plane. The field lines in the central part of the flux rope are shown with magenta lines.
}
\label{Fig:B_mag}
\end{center}
\end{figure*}

\subsection{Anisotropy in the vortex flows} 

The flux rope eruption tends to generate  a vortex ring that is equivalent to the hydrodynamic case. However, the presence of the magnetic field induces an anisotropy that is not observed in the hydrodynamic case. As a result of this, the vortex ring is actually divided in two arcs located around the flanks of the erupting flux rope (see Figure~\ref{Fig:3D-Rot_u}).  
Figure~\ref{Fig:Rot_u-xzview}  shows that the vortex motions are present along the plane almost parallel to the flux rope axis, but they are not present along the plane almost perpendicular to the axis. 

Let's first consider the dynamic in the plane almost parallel  to the axis of the flux rope (Figure~\ref{Fig:Rot_u-xzview1}). In this case the largest component of the flow velocity is approximately perpendicular to the apex  of the cyan/blue magnetic field lines.  Thus the flow  encounters no significant opposition, neither by the magnetic tension that acts along the axis of curvature of the field line itself, nor by the magnetic pressure gradient because the neighboring field lines are also subjected to a similar flow. As a consequence the magnetic field moves following the flow. 

Different field lines move in different ways because they are advected by different portions of the flow field. In particular, while some field lines expand some other field lines located in the returning part of the vortex flows contract because they are advected towards the central part of the active region (Figure~\ref{Fig:Rot_u-xzview1} and Movie~2). This is clearly the case for FL~2 and FL~3 discussed in Section~\ref{Sec:T-Loops}.  

On the plane almost perpendicular  to the axis of the flux rope (Figure~\ref{Fig:Rot_u-xzview2}), the behavior is significantly different. In this case the largest component of the vortex flows would be lying on the plane that contains the field lines and any motion acts against the magnetic tension. Therefore, in this case the magnetic tension force affects the dynamics of the flow, eventually inhibiting it. 

The above result confirms and extends the earlier finding of \cite{For1990}. In their 2D low-$\beta$ simulation the vortex flows develops along a plane that is perpendicular to the axis of the flux rope (because of the 2D geometry this is the only plane where they could develop) and did not affect significantly the evolution of the magnetic field. In our simulation the low-$\beta$ regime is brought to its extreme, i.e., $\beta=0$, and the flows in a plane perpendicular to the flux rope's axis do not even develop. We expect that the behavior  of the solar corona should be some how in between the two cases. 

To summarize, vortex flows are present only on a plane aligned with the axis of the erupting flux rope and localized around the flanks of it,  while they are absent on a plane perpendicular to the flux rope. As a result, magnetic field lines that are located in the top part of the vortices will display expansion and rotation, and contraction if/when they approach the returning part of the vortex-flow. Of course, if a peripheral field line happens to be located in the returning part of the vortices then it will essentially be seen as a contracting one, without displaying the initial expansion.  Furthermore, field lines that are highly inclined and thus closer to the photosphere, are (1) more probable to be located in the return part of the vortex flow (Figure~\ref{Fig:Evol} and Movie~1) and (2) have a higher density and therefore are the ones preferably visible in observations.

\section{Sink flows}
\label{Sec:T-Inflow}

In this Section we describe the inflows towards the current sheet that develop below the magnetic flux rope. The presence of these flows can be deduced from Figure~\ref{Fig:Evol} (top-panels) and accompanying Movie~1 that show the motion of the legs of the overlying field towards the region that lies below the expanding flux rope.

\subsection{Flow sink at the flare's current sheet}

An analysis of Movie~4, shows that during the initial phase of the vortex formation, i.e., between [$164~\tA , 184~\tA$], the flows below the erupting flux rope essentially display an upward directed motion. This is true up to about $t=192~\tA$, when a deflection downwards starts to be visible.

At time $t=204~\tA$, the $\nabla \cdot \mathbf{u}=-0.15$ iso-surface appears and grows in size until the end of the simulation (Figure\ref{Fig:Div_u_3D}a). The snapshot of Movie~4 at $t=204~\tA$ shows that the flow approaches  from both sides of this region, then the flow  splits into upward and downward directed flows. This is true until the end of the simulation, with the extra effect that the flow actually approaches the growing  $\nabla \cdot \mathbf{u}<0$ region at faster and faster speeds both as a function of time and as a function of the distance from the $\nabla \cdot \mathbf{u}<0$ layer. 

A comparison between Figure~\ref{Fig:Div_u_3D}(b) and \ref{Fig:Div_u_3D}(c) shows how the  $\nabla \cdot \mathbf{u}<0$ region (below the most right magenta field line) is spatially correlated with the current sheet that exists below the erupting flux rope. The center of this current sheet coincides with the thinnest region of $\nabla \cdot \mathbf{u}<0$. This  region becomes broader at the two extremities of the current sheet.

Figure~\ref{Fig:1D-Rot_u} (bottom panels) shows a one-dimensional cut along a direction roughly perpendicular to the planar $\nabla \cdot \mathbf{u}<0$ region. This Figure shows that (1) the flow accelerates towards the current sheet, (2) the flow divergence slowly becomes more and more negative while the simulation develops (during tens of Alfv\'{e}n times), and (3) the flow approaches this region faster and faster while the time passes.

\subsection{Physical origin of the sink-flow}
\label{Sec:T-curr}

In order to test if the observed flow resembles the flow around a hydrodynamic sink, we consider the flow generated by a two-dimensional, planar source of $\nabla \cdot \mathbf{u}$. We choose  this kind of source because the thickness of the iso-surface of $\nabla \cdot \mathbf{u}<0$ is much smaller than its other two linear dimensions (Figure~\ref{Fig:Div_u_3D}a and Movie~4 at $t=244~\tA$).  The theoretical flow profile along a direction orthogonal to center of the planar sink (i.e., along $x$) can be computed by using an analogy with the electric field (see Appendix~1 for the derivation) and it is:
\begin{equation}
u(x) = C~\text{atan}~\frac{L/W}{2~(x/W) \sqrt{4~(x/W)^2+(L/W)^2+1}},
\label{Eq_ef3}
\end{equation}
where $C=(\delta x/2\pi)~\text{min}[\nabla \cdot \mathbf{u}]$ is a parameter related to the thickness of the (almost) planar sink. 

The parameters $\delta x, W, L$ can be derived from the simulation as follows. We approximated the profile of $\nabla \cdot \mathbf{u}$ along $x$ to a triangle of base $\delta x$. For the purpose of this analysis we assume two different possible basis for this triangle: one where $\nabla \cdot \mathbf{u}=0$, i.e., $ \delta x =0.11$, and one just outside the two maxima of $\nabla \cdot \mathbf{u}$ (see Figure~\ref{Fig:1D-Rot_u}(i)), i.e., $\delta x=0.18$. To determine the parameters $W$ and $L$, we consider an iso-surface of $\nabla \cdot \mathbf{u}$ equal to its average value withing the current sheet, i.e., $\nabla \cdot \mathbf{u}=(\text{min}[\nabla \cdot \mathbf{u}]+\text{max}[\nabla \cdot \mathbf{u}])/2$. We find $W=1.3$ and $L=2.3$.  
The theoretical model reproduces the simulated flow reasonably well (see Figure~\ref{Fig:1D_analyt}), suggesting that the flow is indeed  compatible with a sink-flow. 

To clarify the physical origin of this sink-like flow, we studied the evolution of the Lorentz force and of its decomposition in magnetic pressure gradient and magnetic tension terms across the current sheet.  Figure~\ref{Fig:1D-Rot_u}(a-c) shows that a double-peaked Lorentz force directed towards the current sheet itself exists at the edges of the current sheet (maximum in the current density). This is the result of an imbalance between the magnetic pressure gradient and  the magnetic tension terms (panels (d-f)).
 
If the magnetic field at the current sheet is reconnected fast enough the inward-directed magnetic pressure gradient term is maintained.  This leads to a quasi-stationary situation where a $\nabla \cdot \mathbf{u}<0$ region is sustained by the magnetic pressure deficit within the current sheet itself.

Eventually, a current-sheet collapse  occurs in our model. It halts the simulation. However, the collapse takes several tens of Alfv\'{e}n time units of the simulation, which correspond to hundreds of Alfv\'{e}n times across the thickness of the modeled current sheet. This relatively slow evolution of the sink (i.e., the $\nabla \cdot \mathbf{u}<0$ layer) is the property of the system that permits to establish, through the launching of quickly-propagating Alfv\'{e}n waves away from the sink \citep[as in][]{ForbesAlone1982}, a large-scale sink-flow that can quasi-statically adjust to the local conditions met within the current sheet at a given time.
So our model implies that the large-scale coronal inflow that brings distant magnetic field lines towards the reconnection site is controlled by the magnitude of the negative flow divergence inside the current sheet which, in our zero-$\beta$ assumption, is directly due to the double-peaked Lorentz force that is primarily caused by the local magnetic depression within the sheet.

This interesting result implies that, although (the flare) reconnection (and therefore energy release) does occur in our model, it is not directly responsible for the inflow. And it also predicts that finite-$\beta$ effects should  influence these inflows. Indeed, the development of a local over-pressure within the current sheet should diminish the magnitude of the local sink, hence should make the large-scale inflows slower than in the zero-$\beta$ case. 

An alternative possible explanation for the reported inflows towards the current sheet could be that the large scale pressure drop left behind the erupting flux rope  forces the legs of the line-tied envelope towards the current sheet, resulting in the observed inflows.  However, there is essentially no Lorentz force outside the current sheet, apart from at late times when the flux rope legs start to move inward. Also the modeled flow is remarkably compatible with the analytic solution around a sink region that has the same characteristics as the one in our simulation.
Moreover, the presence of a Lorentz force directed towards the current sheet does not depend on the flux rope eruption itself because the same occurs during the relaxation of quasi-force free fields \citep[e.g.][ Sect.~3.3]{Aul2006}. Then, we argue that it is the sink-flow generated at the current sheet that drives the modeled  loops (green/gray field lines in Figure~\ref{Fig:Evol}) towards the reconnection region, rather than the large scale pressure drop left by the erupting flux rope.

\section{Summary and discussion}
\label{Sect:Discussion}

In this paper we investigated the hydro-magnetic mechanisms at the 
origin of recently-observed flare-related active-region loop contractions, 
and of well-known coronal-inflows towards flare current-sheets. This 
study was conducted through the analysis of a three-dimensional line-tied 
visco-resistive MHD simulation of an eruptive flare, achieved with the 
OHM-MPI code. The modeled eruption was triggered by the onset of the torus 
instability of a weakly-twisted flux rope, which was previously formed 
by flux cancellation \citep{Zuc2015}. A key property of this 
simulation is that the corona was treated as a zero-$\beta$ plasma, so 
that thermal pressure and gravity were neglected and the only forces at 
work were magnetic pressure and tension. 

Typical observations as well as a first visual inspection of our model 
tend to suggest that loop contractions would be consistent with a mere 
relaxation of previously expanding loops, which would eventually shrink 
towards the place where the pre-erupting current-carrying field was 
located before the flare, and where the flare reconnection eventually 
occurs. In principle, such a relaxation would be natural since our model 
clearly shows that both the evacuation of the flux rope as well as the 
reconnection do lead to a magnetic energy density decrease in this region 
(see Figure~\ref{Fig:B_mag}), so the related inward magnetic-pressure gradient 
may induce a global implosion. This interpretation was already put forward 
with observational analyses \citep[e.g.][]{Russell15,Wang2016} building 
upon a previously proposed conjecture linking magnetic energy and coronal 
implosions \citep{Hudson00}.

Instead, the model reveals that coronal loops are advected by a pair 
of vortices that develop on both sides of the flux rope legs and that 
eventually reach an active-region scale in a characteristic Alfv\'{e}n
 time. Their rotational motions do not imply a volume decrease, and they 
are unrelated with the magnetic energy decrease that develops behind 
the erupting flux rope. The flows are firstly accelerated by a magnetic 
tension force that develops in the wake of the Alfv\'{e}n wave (which would be 
magneto-acoustic in a finite-$\beta$ regime) that is launched from the 
flux rope as soon as its starts moving. This wave 
evacuates away the compressible component of the flow, leaving behind two vorticity arcs located on the flanks of the flux rope legs. 

This behaviour is reminiscent of experimental generation of vortex rings 
when a fast flow of liquid is injected from a hole into a tank \citep{Gle1988,Sul2008}, 
of modeled high-$\beta$ vortices that form around flux ropes that move 
in the solar convection zone \citep{Emo1998,Jouve09}, and more generally of modeled vortex 
pairs that form around a moving fluid element \citep[][ Figures 
9.7-9.8 pages 367-368]{Belmont2013}. 

Some differences exist between our model and these purely hydrodynamical or high-$\beta$ examples, as
follows. 

In our model, the centrifugal acceleration of the flows is not (and cannot 
be) balanced by thermal pressure. It is rather and mostly ensured by magnetic 
tension, although a magnetic pressure gradient that points away from the 
flux rope also contributes on the side of the vortex that is closest to 
the flux rope. Also, it is not a full vortex ring that forms around the 
eruption, but rather two vorticity arcs. The reason is that the magnetic 
field in the legs of the arcades that overlay the flux rope prevents 
rotational motions to develop in a plane perpendicular to the flux rope's axis 
with  our zero-$\beta$ assumption. 2D finite-$\beta$ simulations still 
show that vorticity can develop in this direction too, but with seldom 
field line deformation \citep{For1990,Wang2009}. Finally, our modeled 
vortices are not left to evolve on their own, once formed. Instead 
they gradually strengthen and move away from each other. The reason 
is that the flux rope acceleration and expansion generate new magnetic 
pressure gradients, which accelerate the vortex flows, and which lead 
them to push against the flux rope legs as their increasing centrifugal 
acceleration tends to make them grow. 

In addition to the vortices and loop contractions, we analyzed the nature of the coronal 
inflows towards the flare current-sheet. We found that those are 
consistent with a sink flow, perpendicular to the flux rope axis and 
to the vortex flows. The relatively good match found between the simulation 
and an idealized analytical model of a sink flow suggests that the inflows 
are not due the large-scale depression left behind the flux rope as it 
erupts away from the active region (see the large-scale color change from 
dark to bright orange at low altitudes in Figure~\ref{Fig:B_mag}), but instead it is due to the very narrow depression that develops within the elongated current 
sheet (see white region in Figure~\ref{Fig:B_mag}b).

The origin of this sink is merely due to the change in sign of the vertical 
component of the magnetic field across the current sheet. This naturally 
produces a local minimum in magnetic pressure inside the sheet. This local 
depression can cause the fast collapse of current sheets, when neither 
of  the internal thermal pressure or the magnetic pressure from another 
component are strong 
enough to halt the inflows  driven by this local magnetic-pressure deficit \citep{ForbesAlone1982}. 

Our theoretical results on loop contractions and coronal inflows 
thus far have only been analyzed in one simulation. So they will have 
to be investigated in other ones, to test whether or not they are  
specific to our physical and numerical settings. However, as such behavior exist also in the hydrodynamic case, 
it should in principle be unrelated to the exact topology of the pre-erupting 
current-carrying field, nor with the eruption trigger mechanism. 
So it is arguable that they should be present in general. But this 
will have to be checked against other solar eruption models that are 
neither based on flux ropes nor on the torus instability. 

In addition, our results imply several observational predictions that 
should be tested in the future. In particular the sink flow interpretation 
predicts a flow magnitude that asymptotically scales inversely proportional 
to the square of the distance from the current sheet (albeit for the 
possible role of magnetic forces from distant loops that could brake 
them). Also since these sink flows are a property of current sheets, the model 
predicts that they should develop both in confined and eruptive flares. On the other side, 
the vortex flows are caused by the flux rope eruption, so  according to the model they should 
not be observed in confined flares, but only in eruptive flares and filament 
eruptions. However, since these vortices are merely initiated by the early 
displacement of current-carrying fields, they should also occur in 
failed eruptions. Finally, vortices are mere rotational motions. So 
they are neither associated with a volume decrease nor with a global 
coronal expansion followed by an implosion sequence. So both expanding 
and contracting loops should be observed co-temporally. 

\section{Appendix 1}

In Section~\ref{Sec:T-curr} we discussed how the flow of the simulation could be interpreted as the flow due to the presence of an hydrodynamic sink. To show this, we use the analogy between a sink flow and an electric field. Let us consider the electric field in a point $P(x,0,0)$ at a distance $x$ from the center of a 2D planar distribution of charge located on the $yz$-plane such that its surface charge density is $\sigma$ within the region $[ -W/2 \leq y \leq W/2, -L/2 \leq z \leq L/2]$, and zero outside. 

The $x-$component of the electric field at the point $P$ is then given by \citep{durand1964}: 
\begin{align}
 \nonumber E_{x}(x) &= \frac{\sigma}{\pi \epsilon_0}~\text{atan}~\frac{L/W}{2~(x/W) \sqrt{4~(x/W)^2+(L/W)^2+1}} \\
&= \frac{\sigma}{\pi \epsilon_0}~ f(x, L, W)
\end{align}
If the charge distribution is not actually planar, but has a given thickness $\delta x \ll [L,W]$ we have:
\begin{align}
\nonumber E_{x}(x)&= \frac{f(x, L, W)}{\pi}~\int_{-\delta x/2}^{\delta x/2} \frac{\rho(\xi) }{\epsilon_0} d\xi \\
&= \frac{f(x, L, W)}{\pi}~\int_{-\delta x/2}^{\delta x/2} (\nabla \cdot \mathbf{E})~ d \xi,
\label{Eq_ef2}
\end{align}
where $\xi$ is an integration variable along the direction perpendicular to the charged plane.

Replacing the electric field with the velocity field, Equation~\ref{Eq_ef2} describes the dynamic of the flow along a direction perpendicular to an (almost) planar hydrodynamic source/sink. 

In our simulation at $t=244~\tA$  and along the selected line, $\nabla \cdot \mathbf{u}$ reaches a minimum value of $\sim -2.6$ and the thickness of the region where $\nabla \cdot \mathbf{u}=0$ is $\delta x \approx 0.11$. However, as shown in Figure~\ref{Fig:1D-Rot_u}(i), $\nabla \cdot \mathbf{u}$ has two local maxima very close to the region of negative divergence.  The  distance between the two closest grid points  such that the two maxima are included within them  is  $ \delta x \approx 0.18$. If we assume a triangular profile for the distribution of $\nabla \cdot \mathbf{u}$ along $x$, the integral in Equation~(\ref{Eq_ef2}) can be evaluated as the area of a triangular surface of base $\delta x~$ and height min$[\nabla \cdot \mathbf{u}]$ (neglecting the positive $\nabla \cdot \mathbf{u}$ of the local maxima).  With these assumptions Equation~(\ref{Eq_ef2}) becomes: 
\begin{equation}
u(x) = C~\text{atan}~\frac{L/W}{2~(x/W) \sqrt{4~(x/W)^2+(L/W)^2+1}},
\label{Eq_ef4}
\end{equation}
where $C= ( \delta x/2\pi)~\text{min}[\nabla \cdot \mathbf{u}]$. 

Moreover, the system undergoes an asymmetric expansion and the actual flow is the combined result of the expansion of the flux rope and flow towards the current sheet. To account for this  the theoretical profile is shift by a value that is half of the net expansion flow, i.e., $\delta u =-0.009$. This is the analytically velocity profile displayed in Figure~\ref{Fig:1D_analyt}.

\begin{acknowledgements}

The authors wish to thank the  anonymous referee for her/his constructive comments that led to the improvement of the manuscript.  F. P. Z. and G. A.  thank S. Antiochos, R. Grappin and J. Klimchuk for the useful discussions. The work of F. P. Z. was funded by a contract from the AXA Research Fund. F. P. Z. is a Fonds Wetenschappelijk Onderzoek (FWO) research fellow. 
J. D. acknowledges the Grant 17-16447S of the Grant Agency of the Czech Republic, the institutional support RVO:67985815 of the Czech Academy of Sciences, and the hospitality provided by the Observatoire de Paris during his 1-month stay. 
S.A.G. acknowledges the financial support of the DIM ACAV and R\'{e}gion Ile de France. 
This work was granted access to the HPC resources of MesoPSL financed by the R\'{e}gion Ile de France and the project Equip@Meso (reference ANR-10-EQPX-29-01) of the programme Investissements d' Avenir supervised by the Agence Nationale pour la Recherche. 
%
%
\end{acknowledgements}



\begin{thebibliography}{66}
\expandafter\ifx\csname natexlab\endcsname\relax\def\natexlab#1{#1}\fi

\bibitem[{{Aulanier} {et~al.}(2012){Aulanier}, {Janvier}, \&
  {Schmieder}}]{Aul2012}
{Aulanier}, G., {Janvier}, M., \& {Schmieder}, B. 2012, \aap, 543, A110

\bibitem[{{Aulanier} {et~al.}(2006){Aulanier}, {Pariat}, {D{\'e}moulin}, \&
  {DeVore}}]{Aul2006}
{Aulanier}, G., {Pariat}, E., {D{\'e}moulin}, P., \& {DeVore}, C.~R. 2006,
  \solphys, 238, 347

\bibitem[{{Belmont} {et~al.}(2013){Belmont}, {Grappin}, {Mottez}, {Pantellini},
  \& {Pelletier}}]{Belmont2013}
{Belmont}, G., {Grappin}, R., {Mottez}, F., {Pantellini}, F., \& {Pelletier},
  G. 2013, {Collisionless Plasmas in Astrophysics} (Wiley-VCH)

\bibitem[{{Carmichael}(1964)}]{Carmichael64}
{Carmichael}, H. 1964, NASA Special Publication, 50, 451

\bibitem[{{Cheng} {et~al.}(2015){Cheng}, {Ding}, \& {Fang}}]{Cheng15}
{Cheng}, X., {Ding}, M.~D., \& {Fang}, C. 2015, \apj, 804, 82

\bibitem[{{Cheng} {et~al.}(2010){Cheng}, {Ding}, \& {Zhang}}]{Cheng10}
{Cheng}, X., {Ding}, M.~D., \& {Zhang}, J. 2010, \apj, 712, 1302

\bibitem[{{Cheng} {et~al.}(2014{\natexlab{a}}){Cheng}, {Ding}, {Zhang},
  {Srivastava}, {Guo}, {Chen}, \& {Sun}}]{Cheng14a}
{Cheng}, X., {Ding}, M.~D., {Zhang}, J., {et~al.} 2014{\natexlab{a}}, \apjl,
  789, L35

\bibitem[{{Cheng} {et~al.}(2014{\natexlab{b}}){Cheng}, {Ding}, {Zhang}, {Sun},
  {Guo}, {Wang}, {Kliem}, \& {Deng}}]{Cheng14b}
{Cheng}, X., {Ding}, M.~D., {Zhang}, J., {et~al.} 2014{\natexlab{b}}, \apj,
  789, 93

\bibitem[{{Cheng} {et~al.}(2013){Cheng}, {Zhang}, {Ding}, {Liu}, \&
  {Poomvises}}]{Cheng13}
{Cheng}, X., {Zhang}, J., {Ding}, M.~D., {Liu}, Y., \& {Poomvises}, W. 2013,
  \apj, 763, 43

\bibitem[{{Dud{\'{\i}}k} {et~al.}(2014){Dud{\'{\i}}k}, {Janvier}, {Aulanier},
  {Del Zanna}, {Karlick{\'y}}, {Mason}, \& {Schmieder}}]{Dudik14a}
{Dud{\'{\i}}k}, J., {Janvier}, M., {Aulanier}, G., {et~al.} 2014, \apj, 784,
  144

\bibitem[{{Dud{\'{\i}}k} {et~al.}(2016){Dud{\'{\i}}k}, {Polito}, {Janvier},
  {Mulay}, {Karlick{\'y}}, {Aulanier}, {Del Zanna}, {Dzif{\v c}{\'a}kov{\'a}},
  {Mason}, \& {Schmieder}}]{Dudik16}
{Dud{\'{\i}}k}, J., {Polito}, V., {Janvier}, M., {et~al.} 2016, \apj, 823, 41

\bibitem[{Durand(1964)}]{durand1964}
Durand, {\'E}. 1964, {\'E}lectrostatique: Les distributions, Electrostatique
  (Masson)

\bibitem[{{Emonet} \& {Moreno-Insertis}(1998)}]{Emo1998}
{Emonet}, T. \& {Moreno-Insertis}, F. 1998, \apj, 492, 804

\bibitem[{{Fletcher} {et~al.}(2011){Fletcher}, {Dennis}, {Hudson}, {Krucker},
  {Phillips}, {Veronig}, {Battaglia}, {Bone}, {Caspi}, {Chen}, {Gallagher},
  {Grigis}, {Ji}, {Liu}, {Milligan}, \& {Temmer}}]{Fle2011}
{Fletcher}, L., {Dennis}, B.~R., {Hudson}, H.~S., {et~al.} 2011, \ssr, 159, 19

\bibitem[{{Forbes}(1982)}]{ForbesAlone1982}
{Forbes}, T.~G. 1982, Journal of Plasma Physics, 27, 491

\bibitem[{{Forbes}(1990)}]{For1990}
{Forbes}, T.~G. 1990, \jgr, 95, 11919

\bibitem[{Glezer(1988)}]{Gle1988}
Glezer, A. 1988, Physics of Fluids, 31, 3532

\bibitem[{{Gosain}(2012)}]{Gosain12}
{Gosain}, S. 2012, \apj, 749, 85

\bibitem[{{Gou} {et~al.}(2016){Gou}, {Liu}, {Wang}, {Liu}, {Zhuang}, {Chen},
  {Zhang}, \& {Liu}}]{Gou16}
{Gou}, T., {Liu}, R., {Wang}, Y., {et~al.} 2016, \apjl, 821, L28

\bibitem[{{Green} \& {Kliem}(2009)}]{Green09}
{Green}, L.~M. \& {Kliem}, B. 2009, \apjl, 700, L83

\bibitem[{{Hannah} \& {Kontar}(2013)}]{Hannah13}
{Hannah}, I.~G. \& {Kontar}, E.~P. 2013, \aap, 553, A10

\bibitem[{{Hirayama}(1974)}]{Hirayama74}
{Hirayama}, T. 1974, \solphys, 34, 323

\bibitem[{{Hudson}(2000)}]{Hudson00}
{Hudson}, H.~S. 2000, \apjl, 531, L75

\bibitem[{{Imada} {et~al.}(2014){Imada}, {Bamba}, \& {Kusano}}]{Imada14}
{Imada}, S., {Bamba}, Y., \& {Kusano}, K. 2014, \pasj, 66, 17

\bibitem[{{Janvier} {et~al.}(2014){Janvier}, {Aulanier}, {Bommier},
  {Schmieder}, {D{\'e}moulin}, \& {Pariat}}]{Janvier14}
{Janvier}, M., {Aulanier}, G., {Bommier}, V., {et~al.} 2014, \apj, 788, 60

\bibitem[{{Janvier} {et~al.}(2015){Janvier}, {Aulanier}, \&
  {D{\'e}moulin}}]{Janvier15}
{Janvier}, M., {Aulanier}, G., \& {D{\'e}moulin}, P. 2015, \solphys, 290, 3425

\bibitem[{{Janvier} {et~al.}(2013){Janvier}, {Aulanier}, {Pariat}, \&
  {D{\'e}moulin}}]{Janvier13}
{Janvier}, M., {Aulanier}, G., {Pariat}, E., \& {D{\'e}moulin}, P. 2013, \aap,
  555, A77

\bibitem[{{Janvier} {et~al.}(2016){Janvier}, {Savcheva}, {Pariat}, {Tassev},
  {Millholland}, {Bommier}, {McCauley}, {McKillop}, \& {Dougan}}]{Janvier16}
{Janvier}, M., {Savcheva}, A., {Pariat}, E., {et~al.} 2016, \aap, 591, 141

\bibitem[{{Jouve} \& {Brun}(2009)}]{Jouve09}
{Jouve}, L. \& {Brun}, A.~S. 2009, \apj, 701, 1300

\bibitem[{{Kallunki} \& {Pohjolainen}(2012)}]{Kallunki12}
{Kallunki}, J. \& {Pohjolainen}, S. 2012, \solphys, 280, 491

\bibitem[{{Khan} \& {Aurass}(2006)}]{Khan06a}
{Khan}, J.~I. \& {Aurass}, H. 2006, \aap, 457, 319

\bibitem[{{Khan} {et~al.}(2006){Khan}, {Fletcher}, \& {Nitta}}]{Khan06b}
{Khan}, J.~I., {Fletcher}, L., \& {Nitta}, N.~V. 2006, \aap, 453, 335

\bibitem[{{Kopp} \& {Pneuman}(1976)}]{Kopp76}
{Kopp}, R.~A. \& {Pneuman}, G.~W. 1976, \solphys, 50, 85

\bibitem[{{Kushwaha} {et~al.}(2015){Kushwaha}, {Joshi}, {Veronig}, \&
  {Moon}}]{Kushwaha15}
{Kushwaha}, U., {Joshi}, B., {Veronig}, A.~M., \& {Moon}, Y.-J. 2015, \apj,
  807, 101

\bibitem[{{Li} \& {Zhang}(2014)}]{Li14}
{Li}, T. \& {Zhang}, J. 2014, \apjl, 791, L13

\bibitem[{{Li} \& {Zhang}(2015)}]{Li15}
{Li}, T. \& {Zhang}, J. 2015, \apjl, 804, L8

\bibitem[{{Liu} {et~al.}(2010){Liu}, {Lee}, {Wang}, {Stenborg}, {Liu}, \&
  {Wang}}]{Liu10b}
{Liu}, R., {Lee}, J., {Wang}, T., {et~al.} 2010, \apjl, 723, L28

\bibitem[{{Liu} {et~al.}(2012){Liu}, {Liu}, {T{\"o}r{\"o}k}, {Wang}, \&
  {Wang}}]{Liu12a}
{Liu}, R., {Liu}, C., {T{\"o}r{\"o}k}, T., {Wang}, Y., \& {Wang}, H. 2012,
  \apj, 757, 150

\bibitem[{{Liu} \& {Wang}(2009)}]{Liu09b}
{Liu}, R. \& {Wang}, H. 2009, \apjl, 703, L23

\bibitem[{{Liu} \& {Wang}(2010)}]{Liu10a}
{Liu}, R. \& {Wang}, H. 2010, \apjl, 714, L41

\bibitem[{{Moore} {et~al.}(2001){Moore}, {Sterling}, {Hudson}, \&
  {Lemen}}]{Moore2001}
{Moore}, R.~L., {Sterling}, A.~C., {Hudson}, H.~S., \& {Lemen}, J.~R. 2001,
  \apj, 552, 833

\bibitem[{{Patsourakos} {et~al.}(2013){Patsourakos}, {Vourlidas}, \&
  {Stenborg}}]{Patsourakos13}
{Patsourakos}, S., {Vourlidas}, A., \& {Stenborg}, G. 2013, \apj, 764, 125

\bibitem[{{Petrie}(2016)}]{Petrie16}
{Petrie}, G.~J.~D. 2016, \solphys, 291, 791

\bibitem[{{Russell} {et~al.}(2015){Russell}, {Sim{\~o}es}, \&
  {Fletcher}}]{Russell15}
{Russell}, A.~J.~B., {Sim{\~o}es}, P.~J.~A., \& {Fletcher}, L. 2015, \aap, 581,
  A8

\bibitem[{{Savage} {et~al.}(2012){Savage}, {McKenzie}, \& {Reeves}}]{Savage12}
{Savage}, S.~L., {McKenzie}, D.~E., \& {Reeves}, K.~K. 2012, \apjl, 747, L40

\bibitem[{{Savcheva} {et~al.}(2016){Savcheva}, {Pariat}, {McKillop},
  {McCauley}, {Hanson}, {Su}, \& {DeLuca}}]{Sav2016}
{Savcheva}, A., {Pariat}, E., {McKillop}, S., {et~al.} 2016, \apj, 817, 43

\bibitem[{{Savcheva} {et~al.}(2015){Savcheva}, {Pariat}, {McKillop},
  {McCauley}, {Hanson}, {Su}, {Werner}, \& {DeLuca}}]{Sav2015}
{Savcheva}, A., {Pariat}, E., {McKillop}, S., {et~al.} 2015, \apj, 810, 96

\bibitem[{{Savcheva} {et~al.}(2012{\natexlab{a}}){Savcheva}, {Pariat}, {van
  Ballegooijen}, {Aulanier}, \& {DeLuca}}]{Savcheva12a}
{Savcheva}, A., {Pariat}, E., {van Ballegooijen}, A., {Aulanier}, G., \&
  {DeLuca}, E. 2012{\natexlab{a}}, \apj, 750, 15

\bibitem[{{Savcheva} {et~al.}(2012{\natexlab{b}}){Savcheva}, {Green}, {van
  Ballegooijen}, \& {DeLuca}}]{Savcheva12b}
{Savcheva}, A.~S., {Green}, L.~M., {van Ballegooijen}, A.~A., \& {DeLuca},
  E.~E. 2012{\natexlab{b}}, \apj, 759, 105

\bibitem[{{Savcheva} {et~al.}(2014){Savcheva}, {McKillop}, {McCauley},
  {Hanson}, \& {DeLuca}}]{Savcheva14}
{Savcheva}, A.~S., {McKillop}, S.~C., {McCauley}, P.~I., {Hanson}, E.~M., \&
  {DeLuca}, E.~E. 2014, \solphys, 289, 3297

\bibitem[{{Schmieder} {et~al.}(2015){Schmieder}, {Aulanier}, \& {Vr{\v
  s}nak}}]{Sch2015}
{Schmieder}, B., {Aulanier}, G., \& {Vr{\v s}nak}, B. 2015, \solphys, 290, 3457

\bibitem[{{Shen} {et~al.}(2014){Shen}, {Zhou}, {Ji}, {Wiegelmann}, {Inhester},
  \& {Feng}}]{Shen14}
{Shen}, J., {Zhou}, T., {Ji}, H., {et~al.} 2014, \apj, 791, 83

\bibitem[{{Shibata} {et~al.}(1995){Shibata}, {Masuda}, {Shimojo}, {Hara},
  {Yokoyama}, {Tsuneta}, {Kosugi}, \& {Ogawara}}]{Shi1995}
{Shibata}, K., {Masuda}, S., {Shimojo}, M., {et~al.} 1995, \apjl, 451, L83

\bibitem[{{Sim{\~o}es} {et~al.}(2013){Sim{\~o}es}, {Fletcher}, {Hudson}, \&
  {Russell}}]{Simoes13a}
{Sim{\~o}es}, P.~J.~A., {Fletcher}, L., {Hudson}, H.~S., \& {Russell}, A.~J.~B.
  2013, \apj, 777, 152

\bibitem[{{Sturrock}(1966)}]{Sturrock66}
{Sturrock}, P.~A. 1966, \nat, 211, 695

\bibitem[{{Su} {et~al.}(2013){Su}, {Veronig}, {Holman}, {Dennis}, {Wang},
  {Temmer}, \& {Gan}}]{Su13}
{Su}, Y., {Veronig}, A.~M., {Holman}, G.~D., {et~al.} 2013, Nature Physics, 9,
  489

\bibitem[{Sullivan {et~al.}(2008)Sullivan, Niemela, Hershberger, Bolster, \&
  Donnelly}]{Sul2008}
Sullivan, I.~S., Niemela, J.~J., Hershberger, R.~E., Bolster, D., \& Donnelly,
  R.~J. 2008, Journal of Fluid Mechanics, 609, 319

\bibitem[{{Sun} {et~al.}(2012){Sun}, {Hoeksema}, {Liu}, {Wiegelmann},
  {Hayashi}, {Chen}, \& {Thalmann}}]{Sun12}
{Sun}, X., {Hoeksema}, J.~T., {Liu}, Y., {et~al.} 2012, \apj, 748, 77

\bibitem[{{Takasao} {et~al.}(2012){Takasao}, {Asai}, {Isobe}, \&
  {Shibata}}]{Takasao12}
{Takasao}, S., {Asai}, A., {Isobe}, H., \& {Shibata}, K. 2012, \apjl, 745, L6

\bibitem[{{Wang} {et~al.}(2009){Wang}, {Shen}, \& {Lin}}]{Wang2009}
{Wang}, H., {Shen}, C., \& {Lin}, J. 2009, \apj, 700, 1716

\bibitem[{{Wang} {et~al.}(2016){Wang}, {Sim{\~o}es}, {Fletcher}, {Thalmann},
  {Hudson}, \& {Hannah}}]{Wang2016}
{Wang}, J., {Sim{\~o}es}, P.~J.~A., {Fletcher}, L., {et~al.} 2016, \apj, 833,
  221

\bibitem[{{Yokoyama} {et~al.}(2001){Yokoyama}, {Akita}, {Morimoto}, {Inoue}, \&
  {Newmark}}]{Yokoyama01}
{Yokoyama}, T., {Akita}, K., {Morimoto}, T., {Inoue}, K., \& {Newmark}, J.
  2001, \apjl, 546, L69

\bibitem[{{Zhang} {et~al.}(2012){Zhang}, {Cheng}, \& {Ding}}]{Zhang12}
{Zhang}, J., {Cheng}, X., \& {Ding}, M.-D. 2012, Nature Communications, 3

\bibitem[{{Zhao} {et~al.}(2016){Zhao}, {Gilchrist}, {Aulanier}, {Schmieder},
  {Pariat}, \& {Li}}]{Zhao2016}
{Zhao}, J., {Gilchrist}, S.~A., {Aulanier}, G., {et~al.} 2016, \apj, 823, 62

\bibitem[{{Zhu} {et~al.}(2016){Zhu}, {Liu}, {Alexander}, \& {McAteer}}]{Zhu16}
{Zhu}, C., {Liu}, R., {Alexander}, D., \& {McAteer}, R.~T.~J. 2016, \apjl, 821,
  L29

\bibitem[{{Zuccarello} {et~al.}(2015){Zuccarello}, {Aulanier}, \&
  {Gilchrist}}]{Zuc2015}
{Zuccarello}, F.~P., {Aulanier}, G., \& {Gilchrist}, S.~A. 2015, \apj, 814, 126

\end{thebibliography}

\end{document}